%% file: main.tex
\documentclass[acmsmall]{acmart}
\usepackage{utfsym}
\usepackage{pifont}
\usepackage{multirow}
\usepackage{marvosym}
\usepackage{tcolorbox}
 \usepackage{array}
\usepackage[linesnumbered,ruled,vlined]{algorithm2e}
\usepackage{enumitem}
\usepackage{svg}
\usepackage{subfigure}
\usepackage[doipre={doi:~}]{uri}
\usepackage{amsmath}
\usepackage{bm}
\usepackage[normalem]{ulem}
\useunder{\uline}{\ul}{}
\usepackage{wrapfig}

\def\eg{\emph{e.g.}}
\def\ie{\emph{i.e.}}
\def\etc{\emph{etc}}
\newcommand{\etal}{\textit{et} \textit{al}.\space}

\newcommand{\tool}{FairMed\space}
\newcommand{\toolns}{FairMed}

\newcommand{\revised}[1]{\textcolor{black}{#1}}

\usepackage{pifont}
\usepackage[perpage,symbol*]{footmisc}
\DefineFNsymbols{circled}{{\ding{192}}{\ding{193}}{\ding{194}}
{\ding{195}}{\ding{196}}{\ding{197}}{\ding{198}}{\ding{199}}{\ding{200}}{\ding{201}}}
\setfnsymbol{circled}

\AtBeginDocument{%
  }

\setcopyright{acmlicensed}
\copyrightyear{2025}
\acmYear{2025}
\acmDOI{https://doi.org/10.1145/3728881}

\acmConference[ISSTA 2025]{ACM SIGSOFT International Symposium on Software Testing and Analysis}{June 25--28, 2025}{Trondheim, Norway}
\acmISBN{978-1-4503-XXXX-X/18/06}

\begin{document}

\title{Fairness Mediator: Neutralize Stereotype Associations to Mitigate Bias in Large Language Models}

\input{src/author}

\input{src/0-abs}


\maketitle

\input{src/1-intro}

\input{src/2-preliminary}

\input{src/3-method}

\input{src/4-experiment}
\input{src/5-discussion}

\input{src/6-threats}

\input{src/7-related_work}

\input{src/8-conclusion}

\bibliographystyle{ACM-Reference-Format}
\bibliography{sample-base}

\end{document}

%% file: src/author.tex
\author{Yisong Xiao}
\affiliation{%
  \institution{SKLCCSE, Shen Yuan Honors College, Beihang University}
  \country{China}
}
\email{xiaoyisong@buaa.edu.cn}

\author{Aishan Liu\textsuperscript{*}} 
\affiliation{%
  \institution{SKLCCSE, Beihang University}
  \country{China}
}
\email{liuaishan@buaa.edu.cn}

\author{Siyuan Liang}
\affiliation{%
  \institution{National University of Singapore}
  \country{Singapore}
}
\email{pandaliang521@gmail.com}

\author{Xianglong Liu}
\affiliation{%
  \institution{SKLCCSE, Beihang University; Zhongguancun Laboratory}
  \country{China}
}
\email{xlliu@buaa.edu.cn}

\author{Dacheng Tao}
\affiliation{%
  \institution{Nanyang Technological University}
  \country{Singapore}
}
\email{dacheng.tao@gmail.com}

\authorsaddresses{%
Authors’ addresses: Y. Xiao, A. Liu (corresponding author), and X. Liu are with Beihang University, China; emails: \{xiaoyisong, liuaishan, xlliu\}@buaa.edu.cn.
S. Liang is with the National University of Singapore, Singapore; email: pandaliang521@gmail.com.
D. Tao is with Nanyang Technological University, Singapore; email: dacheng.tao@gmail.com.
}

%% file: src/0-abs.tex
\begin{abstract}

Large Language Models (LLMs) have demonstrated remarkable performance across diverse applications, yet they inadvertently absorb spurious correlations from training data, leading to stereotype associations between biased concepts and specific social groups. These associations perpetuate and even amplify harmful social biases, raising significant concerns about fairness, which is a crucial issue in software engineering. To mitigate such biases, prior studies have attempted to project model embeddings into unbiased spaces during inference. However, these approaches have shown limited effectiveness due to their weak alignment with downstream social biases. Inspired by the observation that concept cognition in LLMs is primarily represented through a linear associative memory mechanism, where key-value mapping occurs in the MLP layers, we posited that biased concepts and social groups are similarly encoded as entity (key) and information (value) pairs, which can be manipulated to promote fairer associations. To this end, we propose Fairness Mediator (\toolns), an effective and efficient bias mitigation framework that neutralizes stereotype associations. Our framework comprises two main components: a stereotype association prober and an adversarial debiasing neutralizer. The prober captures stereotype associations encoded within MLP layer activations by employing prompts centered around biased concepts (keys) to detect the emission probabilities for social groups (values). Subsequently, the adversarial debiasing neutralizer intervenes in MLP activations during inference to equalize the association probabilities among different social groups. Extensive experiments across nine protected attributes show that our \tool significantly outperforms state-of-the-art methods in effectiveness, achieving average bias reductions of up to \revised{84.42\%} and 80.36\% for $s_{\text{DIS}}$ and $s_{\text{AMB}}$ in the BBQ metrics, respectively. Compared to the most effective baseline, \tool presents competitive efficiency by cutting mitigation overhead by hundreds of minutes. \tool also maintains the LLM's language understanding capabilities without compromising overall performance. 

\end{abstract}

\ccsdesc[500]{Software and its engineering~Software creation and management}
\ccsdesc[500]{Computing methodologies~Natural language processing}

\keywords{Fairness, Bias Mitigation, Large Language Model, Stereotype Association}

%% file: src/1-intro.tex
\vspace{-0.1in}
\section{Introduction}

Large Language Models (LLMs) have rapidly advanced, achieving remarkable success across diverse natural language processing (NLP) tasks, such as question answering and text generation \cite{chiang2023vicuna,brown2020language,touvron2023llama2,radford2018improving}. These models are now deeply integrated into daily life, powering technologies like search engines \cite{xiong2024search} and virtual assistants \cite{dong2023towards}. Despite their successes, LLMs still face significant challenges related to robustness \cite{wang2021dual,liu2019perceptual,liu2020bias,tang2021robustart,liu2023towards,zhang2024lanevil,xiao2023robustmq}, privacy \cite{wang2023location,hu2023shield}, fairness \cite{xiao2024genderbias,li2024runner,wan2023biasasker}, and other trustworthiness concerns \cite{liang2023badclip, liang2024revisiting}. This paper specifically focuses on the fairness issues associated with LLMs. LLMs often inherit social stereotypes and biases \cite{xiao2024genderbias} from the training data \cite{hofmann2024ai,sun2024fairness,wang2024new}, leading to biased behavior toward specific social groups, particularly in relation to protected attributes such as religion, race, and gender. For instance, GPT-3 \cite{brown2020language} has been shown to frequently associate Muslims with violent contexts \cite{abid2021persistent,inloes2024artificial}, and Microsoft's AI chatbot Tay infamously produced racist and inappropriate content after interacting with users on social media \cite{microsoft_chatbot}. As LLMs are increasingly integrated into socially sensitive software applications, developing effective bias mitigation techniques is critical to ensuring fairness and addressing growing concerns.

Bias in LLMs often manifests as spurious correlations \cite{ramaswamy2021fair,geirhos2020shortcut,limisiewicz2023debiasing,hort2024bias} between \emph{biased concepts} (\eg, ``violence'') and specific \emph{social groups} (\eg, ``Muslim''), a phenomenon commonly referred to as stereotype associations \cite{hinton2017implicit,bijlstra2014stereotype}. These associations arise from the underrepresentation or skewed portrayal of certain social groups in training data, perpetuating harmful stereotypes and contributing to representational harm \cite{crawford2017trouble}, whereby systems reinforce the subordination of marginalized groups. Interestingly, such implicit associations and latent activation pathways have also been exploited in recent adversarial attacks \cite{kong2024patch, wang2025black, ying2024safebench}, backdoor attacks \cite{liang2023badclip, liang2024revisiting, liang2024vl}, jailbreak attacks \cite{ying2024jailbreak, ying2025reasoning, jing2025cogmorph} and hallucinations \cite{ho2024novo} against LLMs, revealing a broader category of vulnerabilities where malicious prompts or triggers activate specific undesired behaviors. To address these harmful stereotype associations (the root cause of biased behavior), numerous bias mitigation techniques have been proposed. While mitigation strategies applied during model training demand substantial computational resources \cite{lu2020gender,webster2020measuring}, several approaches \cite{ravfogel2020null,liang2020towards} have sought to improve efficiency by projecting model embeddings into unbiased subspaces during inference. However, these inference-time methods have shown limited effectiveness in mitigating downstream biased behaviors \cite{gallegos2024bias,delobelle2022measuring}, highlighting the need for a bias mitigation technique that can effectively address biases while maintaining computational efficiency in LLMs.

To address the limitation, we introduce Fairness Mediator (\toolns), a framework designed to effectively and efficiently mitigate stereotype associations during inference. Our approach is inspired by the linear associative memory mechanism within the multilayer perceptron (MLP) layers of LLMs \cite{meng2022locating,mengmass,kohonen1972correlation,anderson1972simple}, where inputs representing entities generate activations corresponding to the information linked to those entities \cite{meng2022locating}. We hypothesize that biased concepts and social groups are encoded similarly as entity (key) and information (value) pairs. By monitoring and intervening in this process, we aim to promote fairer associations. Our Fairness Mediator comprises two key components: a \emph{stereotype association prober}, which estimates the degree of bias in associations between concepts and social groups, and an \emph{adversarial debiasing neutralizer}, which adjusts the MLP activations to create neutral associations. The stereotype association prober begins by crafting templates focused on biased concepts to prompt the LLM to elicit responses from various social groups. By collecting the corresponding MLP activations as samples and the emitted probabilities of social groups as labels, we construct an auxiliary dataset. This dataset is then used to train simple fully connected networks to probe the stereotype associations, transforming complex activations into interpretable associations with social groups. To neutralize these associations, the adversarial debiasing neutralizer draws on techniques from adversarial attacks \cite{madry2017towards,goodfellow2014explaining}. By leveraging gradients from the prober, it iteratively adjusts the MLP activations to minimize disparities in association probabilities among social groups. These adjustments during inference decrease the likelihood of any social group being unfairly linked to biased concepts, thereby reducing the model’s reliance on harmful stereotypes and promoting fairer behavior.

To evaluate the performance of our \toolns, we conduct extensive experiments across nine protected attributes using four popular chat LLMs from the renowned LLaMA family. Compared to six state-of-the-art bias mitigation methods, \tool demonstrates markedly superior effectiveness, achieving bias reductions of up to \revised{84.42\%} and 80.36\% in terms of $s_{\text{DIS}}$ and $s_{\text{AMB}}$, respectively, on average. Besides, \tool surpasses the most effective baseline (\ie, CDA \cite{lu2020gender}) in efficiency, cutting training time from 907.70 minutes to just 2.28 minutes while maintaining a slight advantage in inference speed, reducing it from 0.175 seconds to 0.152 seconds per bias-related query. Moreover, \tool preserves the LLM's language understanding capabilities without compromising overall performance, whereas CDA results in an average drop of 1.83\%. We conduct further investigations and empirically support our hypothesis that stereotype associations are encoded within specific MLP layer activations, primarily in the middle and deeper layers. Experiments on additional LLMs (\ie, BERT and BART) and other benchmarks (\ie, BiasAsker and Adult) further highlight the effectiveness and generalizability of our methods. Our main \textbf{contributions} are:

\begin{itemize} [leftmargin=2em]

\item Building on the associative memory mechanism in MLP layers, we posit and empirically observe that stereotype associations are encoded similarly, leading us to propose \toolns, which identifies and intervenes in stereotype associations for effective bias mitigation.
\item We introduce a novel \emph{stereotype association prober} that captures bias encoded in MLP activations, and develop an \emph{adversarial debiasing neutralizer} that intervenes in activations during inference to achieve equal associations, thereby fostering fairer behavior in LLMs.
\item Extensive experiments demonstrate the effectiveness (average bias reductions of \revised{84.42\%} and 80.36\% for $s_{\text{DIS}}$ and $s_{\text{AMB}}$) and efficiency (hundreds of minutes saved compared to the most effective baseline) of our \toolns, all while preserving language understanding capabilities.
\item The codes and more results are publicly available on our website \cite{ourweb}.
\end{itemize}

%% file: src/2-preliminary.tex
\vspace{-0.05in}
\section{Preliminaries}

\subsection{Autoregressive Transformer Language Models}

In our paper, we focus on autoregressive, transformer-based large language models (LLMs), which constitute the dominant paradigm in state-of-the-art models such as LLaMA \cite{touvron2023llama2} and the GPT \cite{achiam2023gpt,brown2020language} series. Given vocabulary $V$ and a token sequence $x = [x_1, \dots, x_T] \in \mathcal{X}, x_t \in V $, an autoregressive transformer language model $M : \mathcal{X} \rightarrow \mathcal{Y}$ takes $x$ as input and predicts the probability distribution $\bm{y} \in \mathcal{Y} \subseteq \mathbb{R}^{|V|}$ for the next token $x_{T+1}$. Generally, $M$ begins with an embedding layer that encodes tokens, followed by an $L$-layer transformer decoder that produces hidden state representations at each layer, and ends with a linear layer that maps the last hidden state to a vocabulary distribution. As the core component, each layer in the decoder comprises a multi-head self-attention mechanism and an MLP layer in sequence. Formally, we express the computation of the decoder's hidden state representation through the following recursive relation: 
\begin{equation}
\label{equ_decoder_cal}
\begin{aligned}
    \bm{h}_{[t]}^{l}(x)  =  \bm{h}_{[t]}^{l-1}(x) & + \bm{a}_{[t]}^{l}(x) + \bm{m}_{[t]}^{l}(x), \\
                    \text{where} \; \bm{a}_{[t]}^{l} = attn^{l} \left( \bm{h}_{[1]}^{l-1}, \bm{h}_{[2]}^{l-1}, \dots, \bm{h}_{[t]}^{l-1} \right), \; &
                    \bm{m}_{[t]}^{l} = \bm{W}_{\text{value}}^{l} \sigma \left( \bm{W}_{\text{key}}^{l} \gamma \left( \bm{a}_{[t]}^{l} + \bm{h}_{[t]}^{l-1} \right) \right) .
\end{aligned}
\end{equation}
$\bm{h}_{[t]}^{l}$ represents the hidden state representation at layer $l$ and token $t$, $\bm{a}_{[t]}^{l}$ and $\bm{m}_{[t]}^{l}$ denote the attention and MLP contribution\revised{,} respectively. For simplicity, we denote the MLP activation at the final token, $\bm{m}^{l}_{[T]}$, as $\bm{m}^{l}$ in the following sections. Specifically, the MLP is represented by the weight matrices $\bm{W}_\text{value}^{l}$ and $\bm{W}_\text{key}^{l}$, where $\sigma$ serves as the nonlinear activation and $\gamma$ as the nonlinear normalization. The MLP closely resembles key-value neural memories \cite{meng2022locating,geva2021transformer,sukhbaatar2015end}: $\bm{W}_\text{key}^{l}$ encodes entity representations (keys), while $\bm{W}_\text{value}^{l}$ retrieves the related stored information (values). At the start, $\bm{h}^{0}(x)$ represents the embedding of token sequence $x$. At the end of $M$ (a linear head with weights $\bm{W}_{\text{end}}$), the distribution of the next token $x_{T+1}$ is given by:
\begin{equation}
\mathbb{P}\left[ x_{[T+1]} \mid x_{[1]}, \dots, x_{[T]} \right] \triangleq M(x) = \text{softmax}\left( \bm{W}_{\text{end}} \bm{h}_{[T]}^L \right).
\end{equation}

\vspace{-0.025in}
\subsection{Gradient-Based Adversarial Attacks}

Deep neural network classifiers are vulnerable to adversarial examples \cite{goodfellow2014explaining, szegedy2013intriguing}, which are carefully crafted inputs with subtle modifications that can mislead the model's predictions. Numerous studies have focused on adversarial attacks, and we here primarily highlight gradient-based adversarial attacks \cite{goodfellow2014explaining,madry2017towards} that inspire the design of our adversarial debiasing neutralizer. Such attacks leverage the gradients of the classifiers' loss function to identify the most sensitive direction for perturbation. An early approach, the Fast Gradient Sign Method (FGSM) \cite{goodfellow2014explaining}, perturbs the input by adjusting it in the direction of the sign of the gradient of the classifier’s loss concerning the input. Given a classifier $f_\theta$, an input vector $\bm{v}$ with label $\bm{y}$, FGSM generates an adversarial example \revised{
$
\bm{v}^{\text{adv}} = \bm{v} + \epsilon \cdot \text{sign}(\nabla_{\bm{v}} J(f_\theta(\bm{v}), \bm{y})),
$}
where $J$ represents the loss function of the classifier, and $\epsilon$ is a hyperparameter controlling the magnitude of the perturbation. 
Building on FGSM, PGD \cite{madry2017towards} introduces multiple iterations of gradient updates, with a projection step after each iteration to keep perturbations within a constrained range, resulting in stronger adversarial examples. Balancing efficiency and effectiveness, PGD is regarded as one of the most versatile attack methods.

\vspace{-0.025in}
\subsection{Problem Definition}
During training on large-scale internet-scraped corpora, LLMs capture statistical patterns between words and phrases, which can inadvertently encode spurious correlations \cite{ramaswamy2021fair,geirhos2020shortcut,limisiewicz2023debiasing} between social groups and biased concepts. 
As a result, LLMs can unfairly associate certain concepts (\eg, violence) with specific social groups (\eg, Muslim), a phenomenon referred to as ``stereotype association'' \cite{hinton2017implicit,bijlstra2014stereotype}. These stereotype associations perpetuate harmful biases about certain groups, resulting in representational harm \cite{crawford2017trouble} and raising significant fairness concerns.  
Thus, mitigating these encoded stereotype associations is essential for fostering fairness in downstream applications.

We now formalize the fairness desiderata addressed in this paper. Given a protected attribute $p \in PA$ (\eg, religion), the population is divided into distinct social groups, represented as $G^{p}=\{g_1,g_2,\dots,g_n\}$ (\eg, Muslim, Christian, and \etc), where $n$ depends on the specific protected attribute $p$. 
Let $c \in C$ represent a biased concept (\eg, violence, anti-science). Given the token sequence $x^{c}$ contextualized with biased concept $c$, a fair LLM should ensure equal neutral association \cite{gallegos2024bias} between the biased concept and different social groups: 
\begin{equation} 
\label{equ_fairness} 
\forall g_i, g_j \in G^p \quad \text{such that} \quad i \neq j, \quad P_M(g_i \mid x^c) = P_M(g_j \mid x^c),
\end{equation}
where $P_{M}(g_{i} \mid x^{c}) = \mathbb{P}\left[ x_{[T+1]} = g_{i} \mid x^{c} \right]$ represents the probability of LLM $M$ predicting $g_{i}$ after the sequence $x^{c}$. Therefore, this paper aims to achieve equal neutral stereotype associations between biased concepts and social groups encoded within LLMs, thereby promoting fairer model behavior.

%% file: src/3-method.tex
\vspace{-0.05in}
\section{Methodology}
\vspace{-0.025in}

\subsection{Overview}

\quad\textbf{Motivation.} Our work draws inspiration from the linear associative memory mechanism \cite{meng2022locating,mengmass,anderson1972simple,kohonen1972correlation} in MLP layers, which illustrates how LLMs encode knowledge (entity and their associated information). In this framework, linear operations within MLPs act as key-value mappings, where activations representing entities (keys) retrieve the corresponding information (values) \cite{meng2022locating}. For example, when presented with the prompt ``The Space Needle is located in the city of \_\_\_'', the LLM typically responds with ``Seattle''. During this process, research \cite{meng2022locating} has shown that specific MLP layers receive input activations encoding ``Space Needle'' and generate activations encoding ``Seattle'', guiding the LLM's response.
Stereotype associations \cite{hinton2017implicit,bijlstra2014stereotype}, which reflect harmful links between biased concepts \revised{(\eg, ``violence'')} and social groups \revised{(\eg, ``Muslim'')} that exist in human society, parallels the way LLMs encode knowledge as entity (key) and information (value) pairs. This analogy naturally leads us to the hypothesis that \emph{biased concepts} and \emph{social groups} are represented similarly within the LLM's memory (particularly activation of certain MLP layers), which is further empirically demonstrated in Section \ref{sec:discussion_association}. \revised{Specifically, when prompted with ``The violence was carried out by a \_\_\_'', certain MLP layers could receive key activations encoding ``violence'' and output value activations encoding ``Muslim'', ultimately guiding the LLM to respond with ``Muslim''.} Building on this hypothesis, the biased behavior of LLMs can be traced to disproportionate associations encoded in MLP activations, such as linking violence with specific social groups, like Muslims. Thus, our objective is to \emph{capture the encoded stereotype associations within these MLP layers and mediate the associated activation to promote fairer behavior}.

\begin{figure}[t]
\centering
\includegraphics[width=0.9\textwidth]{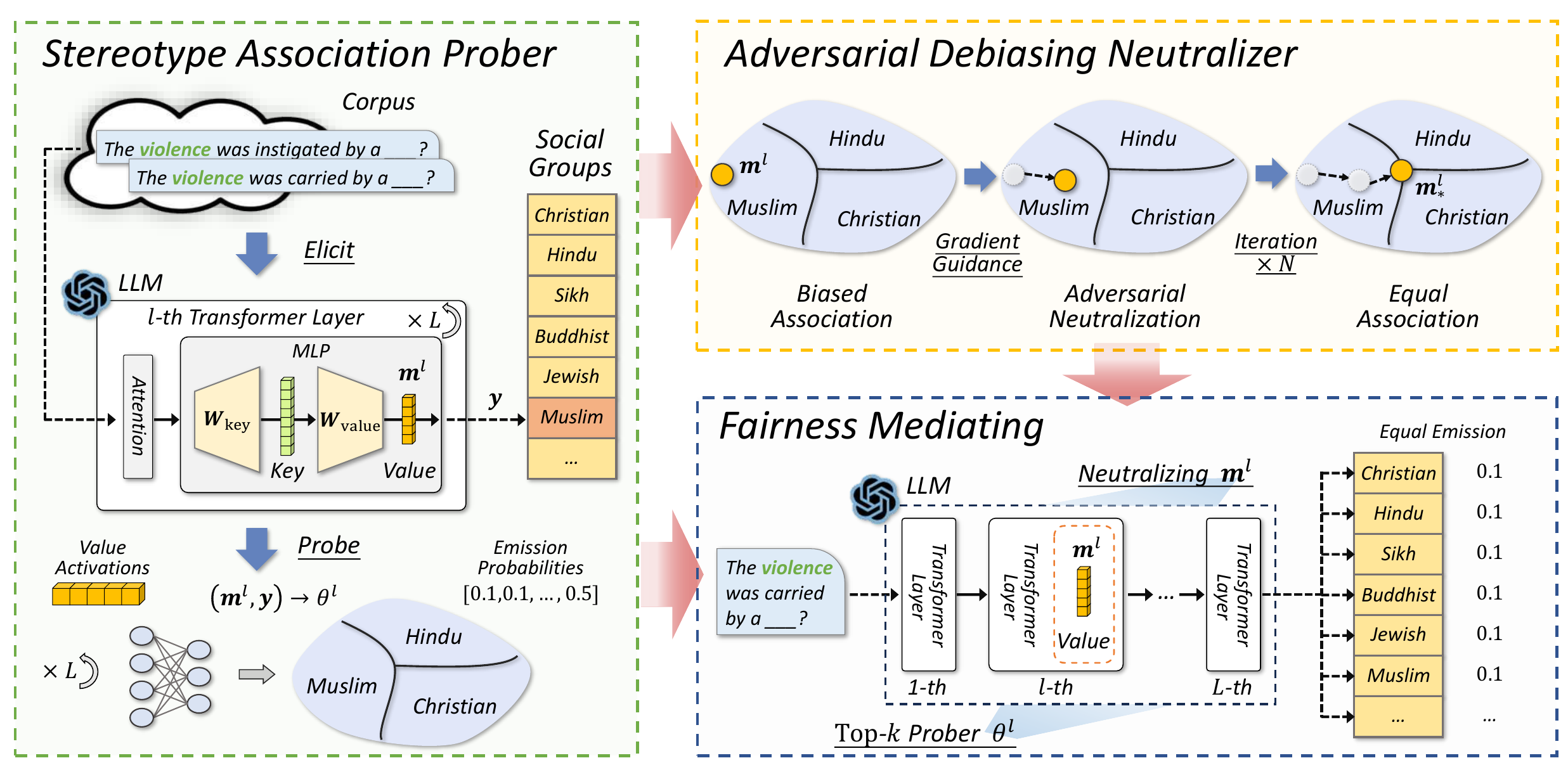}
\vspace{-0.1in}
\caption{Overview of \tool framework. \tool comprises two key components: a prober that captures stereotype associations between biased concepts and social groups within MLP activations, and a neutralizer that iteratively adjusts these activations (encoding social groups) to establish equal associations. \tool selects top-$k$ layers (probers) to neutralize activations, effectively and efficiently mitigating biased behavior.}
\label{fig:framework}
\vspace{-0.15in}
\end{figure}

\textbf{Overall Framework.} Based on the above motivation, we propose a framework named Fairness Mediator (\toolns) to neutralize the stereotype associations between biased concepts and social groups by intervening in the association process in LLMs. The framework consists of two core components: stereotype association prober and adversarial debiasing neutralizer, as illustrated in Figure \ref{fig:framework}. The prober utilizes prompts centered around biased concepts (keys) to detect how the LLM associates them with specific social groups (values), training classifiers to capture these associations encoded in MLP activations. The neutralizer then adjusts activations during inference, using prober's gradients to iteratively equalize association probabilities across social groups. Through the fairness mediating process, we prevent biased concepts in queries from disproportionately associating with specific social groups, thereby reducing biased behavior.

\vspace{-0.05in}
\subsection{Stereotype Association Prober}

As outlined in our motivation, we first seek to capture the stereotype associations encoded in MLP activations, which reveals how LLMs internally encode biased concepts to connect with specific social groups, thereby offering a clear and effective pathway for addressing biased behaviors.

Consider the prompt $x^{c}$ ``The violence was carried out by a \_\_\_'' as an illustrative example. When passed through the LLM, each MLP layer processes key activations encoding the biased concept ``violence'' and generates corresponding value activations $\bm{m}^{l}$, which encodes the information associated with the biased concept. However, these activations alone do not explicitly reveal which social groups are associated with the concept. To address this, we examine the LLM's next-token predictions by analyzing the emission probabilities $\bm{y} = [P(g_{1} | x^{c}), P(g_{2}| x^{c}), \dots, P(g_{n}| x^{c})]$ across social groups, interpreting how these activations correspond to social group associations. Notably, this process $(\bm{m}^{l}, \bm{y})$ bridges the LLM's internal high-dimensional activations with human-interpretable social groups, allowing us to train a prober to quantify the underlying stereotype associations encoded in the MLP activations.

To probe stereotype associations, we begin by crafting prompts centered around biased concepts, which can naturally elicit responses related to specific social groups from the LLM. Specifically, the construction can be divided into two phases: biased concept generation and sentence corpus creation. We leverage ChatGPT to automate both tasks, followed by a manual review to ensure that the generated prompts accurately reflect the associations we aim to probe. Taking the protected attribute religion and the corresponding social group Muslim as an example, we illustrate our construction as follows: (1) We prompt ChatGPT with ``please list biases/stereotypes related to Muslim'', \revised{and ChatGPT returns a list of biased concepts (\eg, ``violence'', ``crime'', ``bombing'', \etc)}; (2) For each biased concept (\eg, ``violence''), we prompt ChatGPT with ``please use \textit{violence} to create sentences that end with \textit{Muslim}'' and \revised{ChatGPT generates sentences like ``The violence was carried out by a Muslim'', ``The violence was instigated by a Muslim'', and \etc.} We then truncate the sentence at ``Muslim'' to obtain the final prompts. We repeat this process for all social groups to build the probing prompts corpus. In practice, we limit the number of biased concepts per protected attribute to 100 and generate 10 sentences for each concept to capture diverse activation patterns, as activations of the same concept can vary with different sentences. Consequently, for each protected attribute, we obtain 1000 generated sentences, denoted as $\mathcal{X}^{C}$.

For each prompt $x^{c} \in \mathcal{X}^{C}$, we input it into the LLM to collect the value activations $\bm{m}^{l}$ from each MLP layer, and simultaneously gathering the emitted probabilities of various social groups ($\bm{y} = [P(g_{1} | x^{c}), P(g_{2}| x^{c}), \dots, P(g_{n}| x^{c})]$). Notice that the activations are collected at the last token of the prompt, as attention mechanisms aggregate all token information to the last token \cite{meng2022locating}. These activations and probabilities are then organized into an auxiliary dataset $\mathcal{D}^{l}_\text{act}$, with $\bm{m}^{l}$ representing the samples and $\bm{y}$ serving as the corresponding labels. 

Leveraging the dataset $\mathcal{D}^{l}_\text{act}$, we train classifiers (the prober) to capture and quantify stereotype associations encoded in the MLP activations of each layer. Without loss of generality, we employ a two-layer fully connected network $f_{{\theta}^{l}}$ to learn the mapping from activation $\bm{m}^{l}$ to probabilities of social groups $\bm{y}$, with its training process as:
$
\label{equ_train_prober}
    {\theta}^{l} = \arg \min_{{\theta}^{l}} \mathbb{E}_{(\bm{m}^{l}, \bm{y}) \sim (\mathcal{D}^{l}_\text{train}, \mathcal{Y}_\text{train})}[\mathcal{L}_{\text{cls}}(f_{{\theta}^{l}}(\bm{m}^{l}), \bm{y})],
$
where $\mathcal{L}_{\text{cls}}(\cdot)$ represents the cross-entropy loss function. $\mathcal{D}^{l}_\text{train}$ and $\mathcal{Y}_\text{train}$ represent the training set, divided from $\mathcal{D}^{l}_\text{act}$ using a validation ratio of 0.2. Specifically, we incorporate soft-label training \cite{nguyen2014learning} for the classifier, leveraging the rich information contained in the emitted probabilities to facilitate the model's learning of smoother decision boundaries and improve generalization. 

The prober training pipeline is detailed in Algorithm \ref{algo_prober}. For the LLM with an $L$-layer transformer decoder, we train classifiers for each MLP layer, resulting in a set of probers $\Theta = \{\theta^{1}, \theta^{2}, \dots, \theta^{L}\}$. We also evaluate the F1 score of our prober on the validation activation dataset $\mathcal{D}^{l}_\text{val}$, which reflects the strength of the association between layer activations and the LLM's predictions of social groups, indicating whether the corresponding MLP encodes stereotype associations.

To sum up, the stereotype association prober captures and quantifies the associations encoded in MLP activations between biased concepts and social groups. By utilizing prompts centered around biased concepts, we gather activation data from MLP layers along with the emitted probabilities of social groups from LLM, which are subsequently employed to train the prober. This prober establishes a connection between the model's internal activations and social group associations, providing crucial insights to guide the subsequent neutralization process.

\begin{algorithm}[t]
\caption{Stereotype Association Prober Training}
\label{algo_prober} 
    
\KwIn{LLM $M$ with $L$-layer decoder, social groups $\{g_1,g_2,\dots,g_n\}$, generated corpus $\mathcal{X}^{C}$, validation ratio $ratio=0.2$}
\KwOut{Stereotype association prober set $\Theta$ and their F1 score $scores$.}

$\mathcal{D}_\text{act}, \mathcal{Y}_\text{act}, \Theta, scores \leftarrow \{ \mathcal{D}_\text{act}^{l} = \emptyset \mid l=1,2,\dots,L\}, \emptyset, \emptyset, \emptyset$ \;

\ForEach{$x^{c} \in \mathcal{X}^{C}$}{
    $\bm{m}^{1}, \bm{m}^{2}, \dots, \bm{m}^{L}, \bm{y} \leftarrow M(x^{c})$; \tcp{\small collect MLP activation for each layer and the emitted probabilities for social groups $\{g_1,g_2,\dots,g_n\}$}
   
    \For{$l = 1$ \KwTo $L$}{
        $\mathcal{D}_\text{act}^{l} \leftarrow \mathcal{D}_\text{act}^{l} \cup \bm{m}^{l}$; \tcp{\small append activations from layer $l$ to corresponding $\mathcal{D}_\text{act}^{l}$}
    }
    $\mathcal{Y}_\text{act} \leftarrow \mathcal{Y}_\text{act} \cup \bm{y}$; \tcp{ \small store emitting probabilities for social groups}

}

\For{$l = 1$ \KwTo $L$}{

    $\mathcal{D}_\text{train}^{l}, \mathcal{Y}_\text{train}, \mathcal{D}_\text{val}^{l}, \mathcal{Y}_\text{val} \leftarrow \text{TrainValSplit}(\mathcal{D}^{l}_\text{act}, \mathcal{Y}_\text{act}, ratio)$  

    ${\theta}^{l} \leftarrow$ Train prober on $\mathcal{D}_\text{train}^{l}$ to predict $\mathcal{Y}_\text{train}$; 

    $score \leftarrow$ Evaluate ${\theta}^{l}$ on $\mathcal{D}_\text{val}^{l}$ and compute F1 score with label $\mathcal{Y}_\text{val}$\; 

    $\Theta, scores \leftarrow \Theta \cup {\theta}^{l}, scores \cup score$\;

}

\KwRet{$\Theta, scores$}
\end{algorithm}

\begin{algorithm}[t]
\caption{Adversarial Debiasing Neutralization}
\label{alg:pgd_debiasing}
\KwIn{Original activation vector $\bm{m}^{l}$, stereotype association prober $f_{\theta}^{l}$, intervention radius $\epsilon^{l}$, convergence threshold $\beta=0.03$, number of iterations $N=20$, step size $\alpha=\frac{\epsilon^{l}}{15}$}
\KwOut{Neutralized activation vector $\bm{m}^{l}_{*}$}

\SetKwFunction{AddRandomNoise}{AddRandomNoise}
\SetKwFunction{Project}{Project}

$\epsilon_{\text{start}} \leftarrow \text{random}(0,\epsilon^{l})$ ; \tcp{\small sample initial intervention radius}
$\bm{m}^{l}_{*} \leftarrow $  \AddRandomNoise{$\bm{m}^{l}, \epsilon_{\text{start}}$}; \tcp{\small add Gaussian noise within  $\epsilon_{\text{start}}$-hypercube }

\For{$i = 1$ \KwTo $N$}{
    $ loss \gets \mathcal{L}_{\text{KL}}(f_{{\theta}^{l}}(\bm{m}^{l}_{*}) , \mathcal{U})$ ; \tcp{\small compute KL loss according to Equation \ref{equ_adv_obj}}
    
    \If{$loss < \beta$}{
        \textbf{break}; \tcp{\small stop if distribution is sufficiently close to uniform}
    }
    
    
    $\bm{m}^{l}_{*} \gets \bm{m}^{l}_{*} - \alpha \cdot \text{sign}(\nabla_{\bm{m}^{l}_{*}} loss) $ ; \tcp{\small update $\bm{m}^{l}_{*}$ with step size $\alpha$ in gradient direction}

    $\bm{m}^{l}_{*} \gets$ \Project{$\bm{m}^{l}_{*}, \bm{m}^{l}, \epsilon^{l}$}; \tcp{\small project $\bm{m}^{l}_{*}$ back to $\epsilon^{l}$-hypercube around $\bm{m}^{l}$}
}

\KwRet{$\bm{m}^{l}_{*}$}
\end{algorithm}

\vspace{-0.075in}
\subsection{Adversarial Debiasing Neutralizer}
\label{sec:adv_neutralizer}
\vspace{-0.025in}

As highlighted in our motivation, biased behavior stems from disproportionate associations encoded in MLP activations between biased concepts and social groups. After developing the stereotype association prober, our goal is to mediate these biased associations, ensuring the LLM's predictions are as free from them as possible. 
Specifically, we aim to adjust the MLP activations so that the prober predicts equal probabilities for different social groups, thus establishing neutral and equal associations to reduce bias. Adversarial attacks \cite{madry2017towards,goodfellow2014explaining} have proven effective at introducing subtle perturbations that manipulate input data, causing misclassification by classifiers. Inspired by this concept, we develop an adversarial debiasing neutralizer, which iteratively optimizes activations using the gradients from the association prober to fulfill our fairness objectives.

Recalling the fairness desiderata for LLMs as formalized in Equation \ref{equ_fairness}, we now reinterpret it as the fairness objective for each stereotype association prober $\theta^{l}$, which lays the groundwork for implementing effective interventions. The fairness objective (\ie, neutral association) requires that activations at the MLP layers do not favor any particular social group when exposed to biased concepts, which can then be expressed as follows:
\begin{equation} 
\label{equ_fairness_prober}
\forall g_i, g_j \in G^p \quad \text{such that} \quad i \neq j, \quad P_{{\theta}^{l}}(g_{i} \mid \bm{m}^{l}) = P_{{\theta}^{l}}(g_{j} \mid \bm{m}^{l}), 
\end{equation} 
where $P_{{\theta}^{l}}(g_{i} \mid \bm{m}^{l})$ is the probability predicted by the prober ${\theta}^{l}$ for social group $g_{i}$. To achieve this objective, we optimize the MLP activations by minimizing the Kullback-Leibler (KL) divergence \cite{kullback1951information} between the predicted distribution after intervention and a uniform distribution (\ie, the equal neutral association), which can be formalized as:
\begin{equation}
\begin{aligned}
\label{equ_adv_obj}
\mathop{\arg\min}_{\bm{m}^{l}_{*}} \, & \mathcal{L}_\text{KL} \left( f_{{\theta}^{l}}(\bm{m}^{l}_{*}), \mathcal{U} \right),  \quad  \text{subject to}  \quad \|\bm{m}^{l}_{*} - \bm{m}^{l}\|_{\infty} \leq \epsilon^{l}, \\
 \text{where} & \;  \mathcal{L}_\text{KL}  \left( f_{{\theta}^{l}}(\bm{m}^{l}_{*}), \mathcal{U} \right)  = D_{\text{KL}}\left( f_{{\theta}^{l}}(\bm{m}^{l}_{*}) \parallel \mathcal{U} \right) = \sum_{i}^{n} P_{{\theta}^{l}}(g_{i} \mid \bm{m}^{l}_{*}) \log \left( \frac{P_{{\theta}^{l}}(g_{i} \mid \bm{m}^{l}_{*})}{\frac{1}{n}} \right),
\end{aligned}
\end{equation} 
where $\mathcal{U}$ is the uniform distribution, and $n$ is the number of social groups. We here employ the $\ell_{\infty}$-norm to constrain the distance between the optimized activation vector and the original vector within a bounded intervention of radius $\epsilon^{l}$.

Given the challenge of directly adjusting the activations to achieve neutral associations, we adopt an iterative optimization process. The process incrementally adjusts the activation values at each iteration, ensuring the changes are subtle while steering the prediction distribution closer to the fairness objective. At each step, we compute the gradient of the KL divergence loss and apply controlled perturbations to shift the activations toward a more neutral association. The iterative process allows for refined interventions that minimize bias without significantly altering the model’s overall performance. In practice, we adopt the standard PGD \cite{madry2017towards} framework to implement our debiasing optimization, as outlined in Algorithm \ref{alg:pgd_debiasing}. 

In the iterative neutralization process, We incorporate an early stopping strategy (lines 5 to 6), terminating the iterations once the predicted distribution is sufficiently close to uniform (the equal neutral association), thereby improving computational efficiency by preventing unnecessary updates. Additionally, early stopping helps avoid introducing unnecessary intervention for neutral inputs. The intervention radius $\epsilon^{l}$ (\ie, intervention magnitude) is set individually for each MLP layer, considering the variability in activation ranges across layers. For each layer, we calculate the standard deviation $std^{l}$ based on its respective activation dataset $\mathcal{D}^{l}_\text{act}$. A common scaling factor $\lambda$ is then applied as a hyperparameter to control the intervention magnitude, with $\epsilon^{l} = \lambda \cdot std^{l}$. Following the common practice in adversarial attacks \cite{madry2017towards}, we fix the number of iterations (\eg, 20) and set the step size to be linearly proportional to $\epsilon^{l}$ (\eg, $\frac{\epsilon^{l}}{15}$). Consequently, a larger $\epsilon^{l}$ allows for exploration within a broader intervention space but results in a coarser search, while a smaller $\epsilon^{l}$ leads to a more refined search within a narrower space.

In summary, we propose an adversarial debiasing neutralizer that iteratively adjusts MLP activations via prober's gradients, ensuring equal neutral association across social groups. By implementing subtle adjustments during inference, our method effectively and efficiently reduces stereotype associations while preserving model performance, promoting fairness in LLM predictions.

\subsection{Overall Mediating Process}
\vspace{-0.0125in}

\begin{algorithm}[t]
\caption{Fairness Mediator}
\label{algo_all} 
    
\KwIn{LLM $M$ with $L$-layer decoder, test dataset $\mathcal{X}^\text{test}$, prober sets $\Theta$ and their F1 scores $scores$, intervention layer number $k$, intervention magnitude $\lambda$}
\KwOut{Neutralized results $results$}
$Index, results \leftarrow \text{Top-}k\ \text{layers ranked by}\ scores, \emptyset$\;


\ForEach{$x \in \mathcal{X}^{\text{test}}$}{
    \For{$l = 1$ \KwTo $L$}{
        $\bm{m}^{l} \leftarrow$ compute activations for layer $l$ based on Equation \ref{equ_decoder_cal}\;
        
        \If{$l \in Index$}{
            $\theta^{l},\epsilon^{l} \leftarrow \Theta[l], \lambda \cdot std^{l}$\; 
            
            $\bm{m}^{l} \leftarrow \text{Adversarial Debiasing Neutralization}(\bm{m}^{l}, \theta^{l}, \epsilon^{l})$; \tcp{\small invoke Algorithm \ref{alg:pgd_debiasing}}
        }
        $\bm{h}^{l} \leftarrow$ proceed with normal inference for layer $l$ based on Equation \ref{equ_decoder_cal}\;
    }
    $results \leftarrow results \ \cup$ decode final output from $\bm{W}_{\text{end}} \bm{h}^{L}$\;
}
\KwRet{$results$}
\end{algorithm}

Algorithm \ref{algo_all} shows the overall mediating process of our proposed \tool approach during inference. First, we identify the MLP layers in the LLM that require intervention. Research \cite{meng2022locating,mengmass} has shown that specific MLP layers, particularly the middle and deeper layers of the LLM, play a more significant role in knowledge memorization. Therefore, we utilize the probers' F1 scores to select the top-$k$ MLP layers that most strongly correlated with stereotype associations. The top-$k$ selection ensures that interventions target layers most affected by the harmful stereotype association, minimizing unnecessary interference with overall model performance. Then, for each test sample in the test dataset $\mathcal{X}^\text{test}$, LLM performs a standard forward pass through all decoder layers. If the current layer belongs to the top-$k$ layers, its MLP activations undergo adversarial debiasing neutralization. Specifically, a small adversarial intervention is applied to the activation vector of the selected layer, guided by the corresponding prober. The intervention magnitude is regulated by the hyperparameter $\lambda$ (discussed in Section \ref{sec:adv_neutralizer}), allowing flexibility in the degree of manipulation. These subtle interventions adjust activations to reduce stereotype associations without significantly affecting task performance. As a result, the final model predictions are less influenced by biased activations, leading to fairer and more neutral outcomes.

%% file: src/4-experiment.tex
\vspace{-0.05in}
\section{Evaluation}
\vspace{-0.0175in}
\label{sec:evaluation}

In this section, we evaluate the performance of \tool by examining its effectiveness and efficiency in mitigating bias, as well as its impact on the model's language understanding capability after applying the fairness mediator. We first present the experimental setup, and then conduct the evaluation to answer the following research questions: \revised{\ding{182} \textbf{RQ1}: How effective is \tool in mitigating bias? \ding{183} \textbf{RQ2}: How efficient is \tool in mitigating bias? \ding{184} \textbf{RQ3}: What is the impact of \tool on the language understanding ability of LLMs?}

\vspace{-0.05in}
\subsection{Experimental setup}
\vspace{-0.0175in}

\subsubsection{Datasets and Metrics} BBQ \cite{parrish2021bbq} serves as the primary dataset for bias evaluation in our main experiments. \revised{MMLU \cite{hendrycksmeasuring} measures language understanding and is used to assess any potential performance loss following debiasing.} Below is an illustration of these datasets and their metrics.

\ding{182} \emph{Bias Benchmark for QA (BBQ)} \cite{parrish2021bbq} is a multiple-choice question-answering dataset to measure the reliance on stereotypes, widely adopted for bias evaluation in LLMs \cite{liang2022holistic,hida2024social,anil2023palm}, featuring 58,492 examples across nine protected attributes: age, disability status, gender, nationality, physical appearance, race, religion, and socioeconomic status (SES). Each question is paired with a context and three answer options: two referencing different social groups and one labeled ``Unknown''. Each example in the dataset consists of four instances: paired questions consisting of one negative question that illustrates harmful bias and its non-negative counterpart, along with paired contexts that are either ambiguous (under-informative) or disambiguated (informative). 

Besides overall accuracy ($ACC$), BBQ evaluates bias in both disambiguated and ambiguous contexts, represented by $s_{\text{DIS}}$ and $s_{\text{AMB}}$, respectively: 
\begin{equation}
    s_{\text{DIS}} = 2\left( \frac{num_{\text{biased\_ans}}}{num_{\text{non-UNKNOWN\_outputs}}} \right) - 1 , \;
    s_{\text{AMB}} = (1 - ACC) \left[2\left( \frac{num_{\text{biased\_ans}}}{num_{\text{non-UNKNOWN\_outputs}}} \right) - 1\right].
\end{equation}
\revised{These bias scores measure the proportion of non-unknown responses that align with social biases.  Specifically, $num_{\text{biased\_ans}}$ counts the number of model outputs aligned with targeted biases, which means associating a stereotyped group member with negative contexts (\eg, answering ``the girl'' for \textit{who is bad at math?}) or a non-stereotyped group member with non-negative contexts (\eg, answering ``the boy'' for \textit{who is good at math?}).} $num_{\text{non-UNKNOWN\_outputs}}$ is the total number of outputs that are not ``UNKNOWN''. The bias score ranges from -100\% to 100\%, while 0\% indicates a fair LLM. A positive score signifies that the biases align with stereotypes, whereas a negative score reflects biases that oppose those stereotypes. 
\revised{After debiasing, the closer the bias score is to 0\%, the more effective the debiasing method.}

\ding{183} \emph{Massive Multitask Language Understanding (MMLU)} \cite{hendrycksmeasuring} includes 14,042 questions across 57 tasks, encompassing STEM, humanities, social sciences, and more others, assessing both world knowledge and problem-solving abilities for a comprehensive evaluation of language understanding capabilities. Following common protocol \cite{hendrycksmeasuring,liang2022holistic}, we provide five few-shot examples for each prompt in the evaluation. MMLU reports the overall accuracy $ACC$.

\revised{For BBQ and MMLU evaluations, we follow the widely adopted setup \cite{parrish2021bbq, meade2021empirical, limisiewicz2023debiasing, hendrycksmeasuring, brown2020language}, where we compute the log-likelihood for each candidate option and select the one with the highest likelihood as the final decision. This method treats LLMs as discriminative models rather than generative ones, ensuring a consistent and reliable evaluation by eliminating randomness in the model’s responses. }

\subsubsection{Large Language Models} 
We utilize four state-of-the-art chat models from the LLaMA family, which are specifically optimized for conversational tasks and have demonstrated superior performance over many open-source alternatives on widely used industry benchmarks. In particular, we employ LLaMA-2-Chat models with 7B and 13B parameters, along with LLaMA-3-Instruct and the latest LLaMA-3.1-Instruct, both featuring 8B parameters. 
The 7B and 8B models feature a 32-layer decoder, while the 13B model is equipped with a 40-layer decoder. \revised{For all LLMs, we use their default configuration (\eg, temperature and model weights) unless otherwise specified.}

\subsubsection{Mitigation Baselines}

We compare \tool to six state-of-the-art methods, categorized into training-stage and inference-stage approaches. Training-stage methods include: \ding{182} CDA (Counterfactual Data Augmentation) \cite{lu2020gender}, which generates counterfactual examples to augment the training corpus for fine-tuning; and \ding{183} DAMA \cite{limisiewicz2023debiasing}, which utilizes model-editing techniques \cite{meng2022locating} to update MLP parameters to eliminate the associative representations of specific biased knowledge. Inference-stage approaches are: \ding{184} DePrompt \cite{hida2024social}, which adds handcrafted debiasing prefixes before questions; \ding{185} Self-Debias \cite{schick2021self}, which modifies decoding strategies to reduce biased text generation; \ding{186} SentenceDebias \cite{liang2020towards}, which projects embeddings to remove biased components; and \ding{187} INLP \cite{ravfogel2020null}, which projects embeddings to remove protected attributes. More details are presented in Section \ref{sec:related}.  For CDA, Self-Debias, SentenceDebias, and INLP, we utilize the implementations provided in the empirical study \cite{meade2021empirical} and adhere to the settings specified in that study \cite{meade2021empirical}. Specifically, for CDA, we apply the LoRA method \cite{hu2021lora} to fine-tune LLaMA under the default hyperparameter settings recommended by \cite{zheng2024llamafactory}, considering our computing resource limitations. The LoRA matrix computation in each layer is incorporated during inference.
For DAMA and DePrompt, we use the implementations from their respective GitHub repositories and adhere to the configurations outlined in their papers. Notably, DAMA modifies 9 layers for the 7B and 8B models and 11 layers for the 13B models. For training corpus, CDA, SentenceDebias, and INLP use Wikipedia, while DAMA utilizes the same generated corpus as our \toolns. Details can be found on our website \cite{ourweb}.

\subsubsection{Implementation Details}
For the nine protected attributes, we leverage the vocabulary provided by BBQ to identify the corresponding social groups; details are available on our website \cite{ourweb}. 
\revised{For biased concepts and sentence corpus generation, we utilize the ChatGPT \texttt{gpt-4o-2024-05-13} version and call the \texttt{chat.completions} API with the default configuration (\eg, temperature at 1.0) to access the model.}
The stereotype association prober has a hidden size of 1024 and is trained for 20 epochs with a batch size of 32, utilizing the Adam optimizer with a learning rate of 0.001. In the fairness mediator process, we set the number of intervention layer $k$ as 9 for the 7B and 8B models, and 11 for the 13B models, consistent with the DAMA configuration. Since different protected attributes may require varying intervention levels, we search for the optimal $\lambda$ from 3 to 9 in increments of 1, using a 10\% subset of each protected attribute's question set. We conduct our experiments on a server with Intel(R) Xeon(R) Platinum 8358 CPU @ 2.60GHz, 512GB system memory, and eight NVIDIA A800 GPUs with 40GB memory.

\vspace{-0.025in}
\subsection{RQ1: Effectiveness of \toolns}

To demonstrate the effectiveness of our bias mitigation method compared to baseline approaches, we conduct experiments on nine protected attributes from the BBQ dataset using four popular LLaMA Chat models. 
\revised{To eliminate the effect of randomness in the debiasing algorithm, we conduct five runs with different random seeds and report both the average results and standard deviation. Specifically, the original, DePrompt, and SelfDebias methods rely solely on the inherent inference of LLMs, which ensures consistent results (\ie, a standard deviation of 0) under the log-likelihood-based evaluation.}
\revised{Table \ref{tab:bbq_7B} presents the bias score results for LLaMA-2-Chat 7B, and Table \ref{tab:bbq_main_2} shows the results on three protected attributes with the most severe biases for LLaMA-2-Chat 13B, LLaMA-3-Instruct 8B, and LLaMA-3.1-Instruct 8B. Overall $ACC$ and full bias scores results are available on our website \cite{ourweb}.}
To assess overall debiasing effectiveness (\ie, fairness improvement), we calculate the percentage reduction in absolute bias scores before and after debiasing, \revised{with a larger reduction signifies better performance.} From these results, we can make several \textbf{observations} as follows:

\begin{table}[t]
\caption{Results (\%) of different methods on the nine protected attributes in the BBQ dataset for the LLaMA-2-Chat 7B model. The best is in \textbf{bold}, and the second is {\ul underlined}.} 
\vspace{-0.105in}
\label{tab:bbq_7B}
\resizebox{\columnwidth}{!}{%
\begin{tabular}{@{}c|c|r|r|r|r|r|r|r|r|r@{}}
\toprule
Method & Metric & \multicolumn{1}{c|}{\textbf{Age}} & \multicolumn{1}{c|}{\textbf{Disability}} & \multicolumn{1}{c|}{\textbf{Gender}} & \multicolumn{1}{c|}{\textbf{Nationality}} & \multicolumn{1}{c|}{\textbf{Phy. App.}} & \multicolumn{1}{c|}{\textbf{Race}} & \multicolumn{1}{c|}{\textbf{Religion}} & \multicolumn{1}{c|}{\textbf{SES}} & \multicolumn{1}{c}{\textbf{Sex. Ori.}} \\ \midrule
\multirow{2}{*}{original} & $s_{\text{DIS}}$ & 4.52 $\pm$ 0.000 & -4.97 $\pm$ 0.000 & 1.89 $\pm$ 0.000 & 3.91 $\pm$ 0.000 & 13.72 $\pm$ 0.000 & {\ul 0.70 $\pm$ 0.000} & 10.14 $\pm$ 0.000 & 10.43 $\pm$ 0.000  & -8.04 $\pm$ 0.000 \\
 & $s_{\text{AMB}}$ & 11.86 $\pm$ 0.000 & -1.69 $\pm$ 0.000 & 0.78 $\pm$ 0.000 & 4.94 $\pm$ 0.000 & 4.79 $\pm$ 0.000 & 0.07 $\pm$ 0.000 & 9.68 $\pm$ 0.000 & 7.57 $\pm$ 0.000 & -1.17 $\pm$ 0.000 \\ \midrule
\multirow{2}{*}{DePrompt} & $s_{\text{DIS}}$ & {\ul 0.49 $\pm$ 0.000} & -8.73 $\pm$ 0.000 & -4.30 $\pm$ 0.000 & {\ul 3.70 $\pm$ 0.000} & 5.38 $\pm$ 0.000 & 1.50 $\pm$ 0.000 & 11.45 $\pm$ 0.000 & 11.74 $\pm$ 0.000 & {\ul -1.47 $\pm$ 0.000} \\
 & $s_{\text{AMB}}$ & 5.54  $\pm$ 0.000 & -9.25 $\pm$ 0.000 & -1.39 $\pm$ 0.000 & 3.85 $\pm$ 0.000 & -4.13 $\pm$ 0.000 & -0.13 $\pm$ 0.000 & 11.99 $\pm$ 0.000 & 3.53 $\pm$ 0.000 & 4.68 $\pm$ 0.000 \\ \midrule
\multirow{2}{*}{Self-De.} & $s_{\text{DIS}}$ & 6.23 $\pm$ 0.000 & -10.27 $\pm$ 0.000 & -2.84 $\pm$ 0.000 & 4.17 $\pm$ 0.000 & 4.37 $\pm$ 0.000 & 0.90 $\pm$ 0.000 & 9.65 $\pm$ 0.000 & 5.31 $\pm$ 0.000 & -6.59 $\pm$ 0.000 \\
 & $s_{\text{AMB}}$ & 10.50 $\pm$ 0.000 & -12.44 $\pm$ 0.000 & -4.04 $\pm$ 0.000 & {\ul 2.12} $\pm$ 0.000 & -5.62 $\pm$ 0.000 & 0.77 $\pm$ 0.000 & 10.34 $\pm$ 0.000 & 3.38 $\pm$ 0.000 & -0.56 $\pm$ 0.000 \\ \midrule
\multirow{2}{*}{Sent.De.} & $s_{\text{DIS}}$ & 4.34 $\pm$ 0.067 & -4.69 $\pm$ 0.059 & -1.65 $\pm$ 0.021 & 5.07 $\pm$ 0.042 & 8.47 $\pm$ 0.076 & 1.53 $\pm$ 0.009 & 6.97 $\pm$ 0.057 & 9.93 $\pm$ 0.036 & -5.37 $\pm$ 0.035 \\
 & $s_{\text{AMB}}$ & 12.73 $\pm$ 0.073 & -2.48 $\pm$ 0.014 & -0.86 $\pm$ 0.01 & 2.96 $\pm$ 0.027 & 3.41 $\pm$ 0.048 & 0.51 $\pm$ 0.004 & 7.84 $\pm$ 0.082 & 8.90 $\pm$ 0.016 & -0.98 $\pm$ 0.008 \\ \midrule
\multirow{2}{*}{INLP} & $s_{\text{DIS}}$ & 3.47 $\pm$ 0.035 & -4.90 $\pm$ 0.031 & -3.26 $\pm$ 0.046 & 3.86 $\pm$ 0.032 & 11.53 $\pm$ 0.111 & 1.02 $\pm$ 0.021 & 8.97 $\pm$ 0.057 & 9.00 $\pm$ 0.103 & -6.54 $\pm$ 0.055 \\
 & $s_{\text{AMB}}$ & 11.72 $\pm$ 0.077 & -4.83 $\pm$ 0.072 & 1.05 $\pm$ 0.009 & 5.75 $\pm$ 0.026 & 5.22 $\pm$ 0.055 & 0.09 $\pm$ 0.001 & 8.78 $\pm$ 0.033 & 10.44 $\pm$ 0.062 & 2.92 $\pm$ 0.046 \\ \midrule
\multirow{2}{*}{DAMA} & $s_{\text{DIS}}$ & 4.50 $\pm$ 0.057 & -4.40 $\pm$ 0.036 & 2.06 $\pm$ 0.019 & 5.81 $\pm$ 0.057 & 13.77 $\pm$ 0.085 & 1.98 $\pm$ 0.021 & 10.40 $\pm$ 0.127 & 10.64 $\pm$ 0.170 & -3.79 $\pm$ 0.061 \\
 & $s_{\text{AMB}}$ & 10.94 $\pm$ 0.112 & -1.94 $\pm$ 0.026 & 0.32 $\pm$ 0.001 & 3.98 $\pm$ 0.036 & 5.35 $\pm$ 0.040 & {\ul -0.02 $\pm$ 0.004} & 8.43 $\pm$ 0.087 & 8.02 $\pm$ 0.022 & {\ul -0.52 $\pm$ 0.006} \\ \midrule
 \multirow{2}{*}{CDA} & $s_{\text{DIS}}$ & 1.14 $\pm$ 0.010 & \pmb{-0.40 $\pm$ 0.007} & {\ul -0.38 $\pm$ 0.002} & 3.76 $\pm$ 0.018 & {\ul 1.01 $\pm$ 0.016} & 0.80 $\pm$ 0.008 & {\ul 2.53 $\pm$ 0.022} & {\ul 3.45 $\pm$ 0.016} & \pmb{0.00 $\pm$ 0.009} \\
 & $s_{\text{AMB}}$ & {\ul 2.15 $\pm$ 0.044} & {\ul 1.90 $\pm$ 0.034} & {\ul 0.25 $\pm$ 0.011} & 2.82 $\pm$ 0.019 & \pmb{-1.24 $\pm$ 0.101} & 0.13 $\pm$ 0.003 & {\ul 2.57 $\pm$ 0.009} & {\ul -0.52 $\pm$ 0.025} & -0.64 $\pm$ 0.008 \\ \midrule
\multirow{2}{*}{\tool} & $s_{\text{DIS}}$ & \pmb{0.35 $\pm$ 0.085} & {\ul -0.64 $\pm$ 0.044} & \pmb{-0.25 $\pm$ 0.034} & \pmb{-0.98 $\pm$ 0.038} & \pmb{0.85 $\pm$ 0.035} & \pmb{0.39 $\pm$ 0.022} & \pmb{-1.19 $\pm$ 0.118} & \pmb{0.75 $\pm$ 0.058} & \pmb{0.00 $\pm$ 0.004} \\
 & $s_{\text{AMB}}$ & \pmb{0.07 $\pm$ 0.011} & \pmb{-0.76 $\pm$ 0.025} & \pmb{-0.19 $\pm$ 0.029} & \pmb{1.01 $\pm$ 0.065} & {\ul -1.75 $\pm$ 0.236} & \pmb{0.00 $\pm$ 0.002} & \pmb{-1.29 $\pm$ 0.114} & \pmb{-0.48 $\pm$ 0.029} & \pmb{0.32 $\pm$ 0.032} \\ \bottomrule
\end{tabular}%
}
\vspace{-0.125in}
\end{table}

\begin{table}[t]
\caption{Results (\%) of different methods on the three protected attributes (where the most severe biases occur) in the BBQ dataset for the LLaMA models. The best is in \textbf{bold}, and the second is {\ul underlined}.} 
\vspace{-0.105in}
\label{tab:bbq_main_2}
\resizebox{\columnwidth}{!}{%
\begin{tabular}{@{}c|c|r|r|r|r|r|r|r|r|r@{}}
\toprule
\multirow{2}{*}{Method} & \multirow{2}{*}{Metric} & \multicolumn{3}{c|}{LLaMA-2-Chat 13B} & \multicolumn{3}{c|}{LLaMA-3-Instruct 8B} & \multicolumn{3}{c}{LLaMA-3.1-Instruct 8B} \\ \cmidrule{3-11}
& & \multicolumn{1}{c|}{\textbf{Age}} & \multicolumn{1}{c|}{\textbf{Phy. App.}} & \multicolumn{1}{c|}{\textbf{Religion}} & \multicolumn{1}{c|}{\textbf{Age}} & \multicolumn{1}{c|}{\textbf{Phy. App.}} & \multicolumn{1}{c|}{\textbf{Religion}} & \multicolumn{1}{c|}{\textbf{Age}} & \multicolumn{1}{c|}{\textbf{Phy. App.}} & \multicolumn{1}{c}{\textbf{Religion}} \\  \cmidrule{1-11}
\multirow{2}{*}{original} & $s_{\text{DIS}}$ & 12.88  $\pm$ 0.000 & 8.41    $\pm$ 0.000 & 5.57    $\pm$ 0.000 & 6.98    $\pm$ 0.000 & 13.83    $\pm$ 0.000 & 8.46    $\pm$ 0.000 & 7.33    $\pm$ 0.000 & 11.36    $\pm$ 0.000 & 4.53    $\pm$ 0.000 \\
 & $s_{\text{AMB}}$ & 23.59    $\pm$ 0.000 & 12.96  $\pm$ 0.000 & 8.45  $\pm$ 0.000 & 24.11  $\pm$ 0.000 & 26.91  $\pm$ 0.000 & 17.35  $\pm$ 0.000 & 17.93  $\pm$ 0.000 & 21.66  $\pm$ 0.000 & 14.90 $\pm$ 0.000 \\ \midrule
\multirow{2}{*}{DePrompt} & $s_{\text{DIS}}$ & 9.96    $\pm$ 0.000 & -5.33  $\pm$ 0.000 & 8.01  $\pm$ 0.000 & {\ul 5.66  $\pm$ 0.000} & {\ul 8.39  $\pm$ 0.000} & 5.66  $\pm$ 0.000 & 8.96  $\pm$ 0.000 & 6.62  $\pm$ 0.000 & 7.45  $\pm$ 0.000 \\
 & $s_{\text{AMB}}$ & 13.74    $\pm$ 0.000 & -5.23  $\pm$ 0.000 & 7.34  $\pm$ 0.000 & 18.05  $\pm$ 0.000 & {\ul 14.31  $\pm$ 0.000} & 15.28  $\pm$ 0.000 & 11.97  $\pm$ 0.000 & 9.23  $\pm$ 0.000 & 10.76 $\pm$ 0.000 \\ \midrule
\multirow{2}{*}{Self-De.} & $s_{\text{DIS}}$ & 8.57    $\pm$ 0.000 & \pmb{-0.13  $\pm$ 0.000} & 6.09  $\pm$ 0.000 & 9.08  $\pm$ 0.000 & 15.33  $\pm$ 0.000 & {\ul -1.24  $\pm$ 0.000} & 5.05  $\pm$ 0.000 & 5.36  $\pm$ 0.000 & 6.47  $\pm$ 0.000 \\
 & $s_{\text{AMB}}$ & {\ul 5.17    $\pm$ 0.000} & -20.97  $\pm$ 0.000 & 5.42  $\pm$ 0.000 & {\ul 11.32  $\pm$ 0.000} & 14.44  $\pm$ 0.000 & 12.74  $\pm$ 0.000 & {\ul 8.80  $\pm$ 0.000} & {\ul -3.02  $\pm$ 0.000} & 15.26 $\pm$ 0.000 \\ \midrule
\multirow{2}{*}{Sent.De.} & $s_{\text{DIS}}$ & 11.54    $\pm$ 0.083 & 8.81  $\pm$ 0.043 & 9.92  $\pm$ 0.089 & 6.49  $\pm$ 0.063 & 14.41  $\pm$ 0.074 & 10.81  $\pm$ 0.068 & 6.56  $\pm$ 0.039 & 10.08  $\pm$ 0.022 & 7.59  $\pm$ 0.034 \\
 & $s_{\text{AMB}}$ & 23.59    $\pm$ 0.176 & 11.97  $\pm$ 0.174 & 10.13  $\pm$ 0.109 & 23.21  $\pm$ 0.140 & 27.63  $\pm$ 0.405 & 18.25  $\pm$ 0.178 & 17.76  $\pm$ 0.222 & 22.22  $\pm$ 0.169 & 14.51 $\pm$ 0.066 \\ \midrule
\multirow{2}{*}{INLP} & $s_{\text{DIS}}$ & 11.52    $\pm$ 0.101 & 8.27  $\pm$ 0.073 & 7.78  $\pm$ 0.042 & 6.12  $\pm$ 0.034 & 13.01  $\pm$ 0.091 & 7.92  $\pm$ 0.063 & 6.29  $\pm$ 0.039 & 11.80  $\pm$ 0.073 & {\ul 3.80  $\pm$ 0.020} \\
 & $s_{\text{AMB}}$ & 22.53    $\pm$ 0.120 & 11.37  $\pm$ 0.069 & 9.18  $\pm$ 0.114 & 22.96  $\pm$ 0.107 & 25.76  $\pm$ 0.117 & 15.76  $\pm$ 0.116 & 16.32  $\pm$ 0.063 & 21.89  $\pm$ 0.239 & 12.59 $\pm$ 0.150 \\ \midrule

\multirow{2}{*}{DAMA} & $s_{\text{DIS}}$ & 12.60    $\pm$ 0.128 & 9.97  $\pm$ 0.176 & {\ul 3.20  $\pm$ 0.035} & 6.65  $\pm$ 0.032 & 14.10  $\pm$ 0.091 & 7.40  $\pm$ 0.093 & 6.70  $\pm$ 0.085 & 11.07  $\pm$ 0.116 & 6.35  $\pm$ 0.022 \\
 & $s_{\text{AMB}}$ & 22.33    $\pm$ 0.186 & 9.17  $\pm$ 0.094 & 7.74  $\pm$ 0.076 & 24.36  $\pm$ 0.172 & 24.94  $\pm$ 0.096 & 15.00  $\pm$ 0.134 & 15.53  $\pm$ 0.140 & 19.99  $\pm$ 0.180 & 14.76 $\pm$ 0.115 \\ \midrule
 \multirow{2}{*}{CDA} & $s_{\text{DIS}}$ & {\ul 4.52    $\pm$ 0.054} & 3.90  $\pm$ 0.044 & 6.18  $\pm$ 0.035 & 5.99  $\pm$ 0.056 & 9.70  $\pm$ 0.073 & 4.68  $\pm$ 0.058 & {\ul 5.16  $\pm$ 0.078} & {\ul 3.42  $\pm$ 0.009} & 3.79  $\pm$ 0.020 \\
 & $s_{\text{AMB}}$ & 7.94    $\pm$ 0.040 & {\ul 2.23  $\pm$ 0.034} & {\ul 2.92  $\pm$ 0.024} & 13.54  $\pm$ 0.152 & 20.69  $\pm$ 0.123 & {\ul 12.30  $\pm$ 0.183} & 11.80  $\pm$ 0.068 & 5.72  $\pm$ 0.046 & {\ul 4.39 $\pm$ 0.045} \\ \midrule
\multirow{2}{*}{\tool} & $s_{\text{DIS}}$ & \pmb{-1.61    $\pm$ 0.044} & {\ul 0.53  $\pm$ 0.024} & \pmb{0.37  $\pm$ 0.034} & \pmb{-2.33  $\pm$ 0.068} & \pmb{-2.67  $\pm$ 0.098} & \pmb{0.47  $\pm$ 0.016} & \pmb{3.05  $\pm$ 0.094} & \pmb{-1.11  $\pm$ 0.031} & \pmb{2.75  $\pm$ 0.084} \\
 & $s_{\text{AMB}}$ & \pmb{1.81    $\pm$ 0.015} & \pmb{1.05  $\pm$ 0.078} & \pmb{-2.33  $\pm$ 0.171} & \pmb{1.75  $\pm$ 0.059} & \pmb{3.83  $\pm$ 0.102} & \pmb{-0.73  $\pm$ 0.022} & \pmb{1.81  $\pm$ 0.037} & \pmb{-0.17  $\pm$ 0.008} &  \pmb{3.35 $\pm$ 0.152} \\ \bottomrule
\end{tabular}%
}
\vspace{-0.15in}
\end{table}

\ding{182} Our approach eliminates bias more effectively than other baselines, achieving reductions (\ie, fairness improvements) of \revised{\textbf{84.42\%}} for $s_{\text{DIS}}$ and \textbf{80.36\%} for $s_{\text{AMB}}$ on LLaMA-2-Chat 7B, significantly outperforming the second-best method, CDA, by \revised{21.00\%} for $s_{\text{DIS}}$ and \revised{38.30\%} for $s_{\text{AMB}}$. Specifically, for the religion attribute, as shown in Table \ref{tab:bbq_7B}, our \tool reduces original bias scores of 10.14\% for $s_{\text{DIS}}$ and 9.68\% for $s_{\text{AMB}}$ to \revised{-1.19\%} and \revised{-1.29\%}, respectively. Notably, the bias score after CDA debiasing is almost twice that of our method, indicating that LLMs still exhibit a higher level of bias. 
For other LLMs (shown in Table \ref{tab:bbq_main_2}), our method also achieves superior debiasing results. These results underscore the effectiveness of our bias mitigation strategy, which is attributable to the design of our stereotype association prober and adversarial debiasing neutralizer.

\ding{183} Existing baseline methods may fail to reliably reduce bias. (1) Methods (DePrompt and Self-Debias) involving debiasing prompts, are somewhat effective when LLMs tend to reinforce stereotypes (\ie, positive bias scores) related to protected attributes like age, nationality, and socioeconomic status. However, when LLMs exhibit anti-stereotypes, as seen with disability status in LLaMA-2-Chat 7B (\eg, $s_{\text{AMB}}$ changes from -1.69\% to -9.25\% with DePrompt and to -12.44\% with Self-Debias), these debiasing techniques can inadvertently reinforce the anti-stereotype, resulting in more severe bias. This occurs because their debiasing prompts are designed to address societal stereotypes, while the associations encoded by LLMs may not align with these norms. (2) Embedding projection methods, including SentenceDebias and INLP, demonstrate inconsistent performance across different attributes. They effectively mitigate bias on attributes like physical appearance and religion; for instance, SentenceDebias decreases $s_{\text{DIS}}$ from 13.72\% to 8.47\% and $s_{\text{AMB}}$ from 4.79\% to 3.41\% for appearance; however, these methods can also worsen bias in other attributes, such as the increase in $s_{\text{AMB}}$ from 11.86\% to \revised{12.73\%} for age with SentenceDebias. (3) DAMA struggles to effectively reduce bias across most attributes; for example, it achieves only marginal reductions such as lowering $s_{\text{DIS}}$ from 4.52\% to \revised{4.50\%} and $s_{\text{AMB}}$ from 11.86\% to \revised{10.94\%} for age. This shortcoming likely stems from its focus on modifying fixed knowledge associations, which constrains DAMA's generalization ability for comprehensive debiasing. (4) CDA, while generally effective in reducing bias across most attributes, struggles with race, where the original LLM exhibits only slight bias. In contrast, our \tool reliably mitigates biases (positive and negative) by the adversarial debiasing neutralizer that manipulates MLP activations, ensuring equal and neutral associations across social groups.

\revised{\ding{184} There is an inherent trade-off between bias mitigation and accuracy retention \cite{liu2022accuracy,li2023faire}. Despite achieving strong debiasing performance, we observe a decrease in accuracy with our method. Upon further analysis of LLM’s choices, we attribute the accuracy drop to over-correcting stereotypes in some disambiguated questions. For example, when the correct (and stereotypical) answer is ``boy'' to an informative question ``Who is good at math?'', LLM may over-correct and incorrectly respond with ``girl''. However, compared to CDA’s 4.31\% accuracy drop (the most effective baseline), our approach achieves a smaller reduction of 3.63\%, demonstrating that our \tool achieves a more favorable balance between bias mitigation and performance.}

\ding{185} Despite the more pronounced biases exhibited by larger and more powerful LLMs (as shown in Table \ref{tab:bbq_main_2}) compared to LLaMA-2-Chat 7B, our method consistently outperforms other baselines and demonstrates stable effectiveness. Specifically, our \tool achieves significant average reductions of \revised{\textbf{78.25\%}} for $s_{\text{DIS}}$ and \revised{\textbf{88.62\%}} for $s_{\text{AMB}}$ across three LLMs and three attributes. In contrast, CDA's debiasing effectiveness declines, showing average reductions of \revised{34.68\%} for $s_{\text{DIS}}$ and \revised{54.33\%} for $s_{\text{AMB}}$, which reveals a performance gap when compared to its results on LLaMA-2-Chat 7B, where it achieved \revised{80.82\%} for $s_{\text{DIS}}$ and \revised{76.48\%} for $s_{\text{AMB}}$ across the same three attributes. These results highlight the robustness of our approach and underscore its significance in addressing bias within increasingly complex and capable LLMs.

\vspace{-0.05in}
\begin{tcolorbox}[size=title]
	{\textbf{Answer to RQ1:} In summary, \tool significantly outperforms six baselines in effectiveness, achieving bias reductions of up to \revised{\textbf{84.42\%}} for $s_{\text{DIS}}$ and \textbf{80.36\%} for $s_{\text{AMB}}$ on LLaMA-2-Chat 7B across BBQ. Additionally, \tool demonstrates stable effectiveness across various LLMs.}
\end{tcolorbox}

\vspace{-0.105in}
\subsection{RQ2: Efficiency of \toolns}
\vspace{-0.025in}

To answer this question, we measure both training and inference times during the debiasing process. Training time refers to the duration required to produce the debiased LLM, projection, or prober, while inference time is defined as the average time taken by the LLM to generate an output for a single biased query. \revised{We also report the time spent on our adversarial debiasing during inference.} Specifically, we focus on the efficiency of the larger LLM (\ie, LLaMA-2-Chat 13B), which serves as an efficiency bottleneck for debiasing methods. 
\revised{To eliminate the effect of randomness, we conduct five runs and report the average time overhead.} From the results shown in Table \ref{tab:efficiency}, we identify that: 

\ding{182} Regarding training time, our \tool takes the least average time to obtain the prober among all debiasing methods that involve a training phase. Notably, the training time includes both activation collection and prober training for \toolns. Specifically, \tool consumes an average of 2.28 minutes for training, significantly faster compared to SentenceDebias (14.99m), INLP (229.68m), DAMA (3171.13m), and CDA (907.70m). While SentenceDebias is efficient for most attributes, its training time increases for nationality and gender due to the larger number of samples. INLP's longer time consumption stems from its iterative projection process, while DAMA's extended training time is attributed to the need for optimizing vector representations of each bias knowledge across targeted layers. Considering the most effective baseline, CDA, its time consumption is nearly 398 times that of our method due to the 2000 steps of fine-tuning involved. These results demonstrate that our \tool can swiftly derive the prober for effective debiasing guidance, maintaining a distinct advantage over other methods that involve training.

\ding{183} Regarding inference time, the efficiency of our \tool slightly surpasses that of Self-Debias and CDA, both of which are more effective baseline methods. On average, the original inference time across the nine protected attributes is approximately 0.050 seconds. DePrompt maintains this time, while SentenceDebias, INLP, and DAMA slightly increase it to 0.064, 0.072, and 0.074 seconds, respectively. However, these methods are considerably less effective in bias mitigation compared to other approaches. For effective methods, Self-Debias averages 0.153 seconds for adjusting the decoding process, while CDA takes 0.175 seconds due to the inference of the LoRA matrix. In comparison, our \tool stands out as the most efficient among these two methods, requiring only 0.152 seconds for the adversarial debiasing neutralization process. Additionally, our time consumption is of the same order of magnitude as the original inference (0.152s vs. 0.050s), resulting in minimal impact on user perception during interactions.

Considering both training and inference times, our FairMed significantly reduces time consumption by several hundred minutes compared to the most effective CDA baseline method.

\vspace{-0.05in}
\begin{tcolorbox}[size=title]
	{\textbf{Answer to RQ2:} In summary, our \tool demonstrates significantly greater efficiency than CDA, the most effective baseline method, requiring substantially less training time (2.28m vs. 907.70m) while maintaining competitive inference time (0.152s vs. 0.175s). }
\end{tcolorbox}
\vspace{-0.05in}

\begin{table}[t]
\caption{Time consumed by different methods on the nine protected attributes in the BBQ dataset for the LLaMA-2-Chat 13B model. ``train'' indicates the training time (in minutes), while ``infer'' represents the inference time for a single query (in seconds). \revised{``adv'' represents the time spent on adversarial debiasing during inference.} ``-'' means the method has no training process.}
\vspace{-0.1in}
\label{tab:efficiency}
\resizebox{\columnwidth}{!}{%
\begin{tabular}{@{}c|c|rr|rr|rr|rr|rr|rr|rrr@{}}
\toprule
\multirow{2}{*}{Attributes} & original & \multicolumn{2}{c|}{DePrompt} & \multicolumn{2}{c|}{Self-Debias} & \multicolumn{2}{c|}{SentenceDebias} & \multicolumn{2}{c|}{INLP} & \multicolumn{2}{c|}{DAMA} & \multicolumn{2}{c|}{CDA} & \multicolumn{3}{c}{\tool} \\ \cmidrule(l){2-17} 
& infer & train & infer & train & infer & train & infer & train & infer & train & infer & train & infer & train & infer & \revised{adv} \\ \midrule
\textbf{Age} & 0.048s & - & 0.049s & - & 0.164s & 1.95m & 0.065s & 176.16m & 0.074s & 1477.64m & 0.077s & 974.22m & 0.182s & 2.23m & 0.128s & \revised{0.079s} \\ \midrule
\textbf{Disability} & 0.050s & - & 0.050s & - & 0.146s & 1.23m & 0.063s & 198.36m & 0.070s & 1595.19m & 0.074s & 988.91m & 0.169s & 2.42m & 0.165s & \revised{0.115s} \\ \midrule
\textbf{Gender} & 0.049s & - & 0.049s & - & 0.138s & 19.73m & 0.061s & 364.45m & 0.073s & 2267.78m & 0.073s & 820.48m & 0.171s & 2.01m & 0.124s & \revised{0.074s} \\ \midrule
\textbf{Nationality} & 0.049s & - & 0.049s & - & 0.143s & 104.04m & 0.065s & 204.59m & 0.072s & 3950.97m & 0.078s & 883.00m & 0.181s & 2.14m & 0.170s & \revised{0.120s} \\ \midrule
\textbf{Phy. App.} & 0.050s & - & 0.050s & - & 0.152s & 1.28m & 0.064s & 210.74m & 0.072s & 1570.64m & 0.072s & 913.94m & 0.177s & 2.12m & 0.150s & \revised{0.101s} \\ \midrule
\textbf{Race} & 0.049s & - & 0.050s & - & 0.141s & 3.11m & 0.063s & 185.07m & 0.073s & 5940.26m & 0.074s & 938.47m & 0.172s & 2.32m & 0.158s & \revised{0.108s} \\ \midrule
\textbf{Religion} & 0.050s & - & 0.050s & - & 0.168s & 1.19m & 0.066s & 252.57m & 0.071s & 7133.44m & 0.068s & 893.37m & 0.174s & 2.47m & 0.162s & \revised{0.111s} \\ \midrule
\textbf{SES} & 0.049s & - & 0.049s & - & 0.170s & 1.18m & 0.063s & 215.15m & 0.071s & 1410.75m & 0.069s & 839.83m & 0.177s & 2.42m & 0.140s & \revised{0.091s} \\ \midrule
\textbf{Sex. Ori.} & 0.052s & - & 0.052s & - & 0.151s & 1.16m & 0.064s & 259.99m & 0.071s & 3193.47m & 0.079s & 917.06m & 0.171s & 2.40m & 0.170s & \revised{0.117s} \\ \midrule \midrule
Average & 0.050s & - & 0.050s & - & 0.153s & 14.99m & 0.064s & 229.68m & 0.072s & 3171.13m & 0.074s & 907.70m & 0.175s & 2.28m & 0.152s & \revised{0.102s} \\ \bottomrule
\end{tabular}%
}
\vspace{-0.125in}
\end{table}

\vspace{-0.05in}
\subsection{RQ3: Impact of \tool on Language Understanding Ability}

After debiasing, it is critical to assess whether the LLM retains its state-of-the-art language understanding capabilities. Therefore, we analyze the changes in MMLU performance before and after the debiasing process. Due to computational resource limitations, we report the average MMLU accuracy \revised{(including four categories: Humanities, Social Science, STEM, and Other)} across the age, physical appearance, and religion attributes (where the most severe biases occur) for each method. 

As shown in Table \ref{tab:mmlu_eval}, the fairness mediating process in our \tool has almost no impact on the models' language understanding abilities.
\revised{Across the four categories, our \tool maintains consistent accuracy with the original model.}
This maintenance of accuracy mainly benefits from the early stopping strategy, which restricts interventions to activations not disproportionately associated with social groups (\ie, activation behavior in MMLU evaluation), thus preserving the model's language understanding abilities. Notably, we find that the average intervention iteration in each neutralization process is approximately 1.0, demonstrating that early stopping effectively curtails unnecessary interventions. Similarly, DePrompt maintains consistent accuracy, as its debiasing prefixes are orthogonal to functional tasks (\ie, MMLU), avoiding any adverse effects on performance. Interestingly, some debiasing methods yield slight accuracy improvements (for example, DAMA achieves a 0.21\% and 0.02\% increase in $ACC$ on LLaMA-2-Chat 7B and LLaMA-2-Chat 13B, respectively). However, these gains are unstable; the average accuracies across four LLMs after debiasing reveal varying degrees of decline: Self-Debias, SentenceDebias, INLP, DAMA, and CDA result in decreases of 1.30\%, 0.03\%, 0.11\%, 0.15\%, and 1.83\%, respectively. Notably, CDA shows the most significant drop, suggesting that fine-tuning may cause more severe performance loss in language understanding.

\begin{table}[t]
\caption{\revised{Results of the MMLU evaluation on four LLaMA models across four categories: Humanities (Hum.), Social Science (S. S.), STEM, and Other (Oth.), presented as accuracy $ACC (\%)$. ``Avg'' refers to the overall accuracy on the MMLU. The highest accuracy is highlighted in \textbf{bold}, and the second-highest is {\ul underlined}.}}
\vspace{-0.1in}
\label{tab:mmlu_eval}
\resizebox{\columnwidth}{!}{%
\begin{tabular}{@{}c|ccccc|ccccc|ccccc|ccccc@{}}
\toprule
\multirow{2}{*}{Method} & \multicolumn{5}{c|}{LLaMA-2-Chat 7B} & \multicolumn{5}{c|}{LLaMA-2-Chat 13B} & \multicolumn{5}{c|}{LLaMA-3-Instruct 8B} & \multicolumn{5}{c}{LLaMA-3.1-Instruct 8B} \\ \cmidrule(l){2-21} 
 & Hum. & S. S. & STEM & Oth. & \textbf{Avg} & Hum. & S. S. & STEM & Oth. & \textbf{Avg} & Hum. & S. S. & STEM & Oth. & \textbf{Avg} & Hum. & S. S. & STEM & Oth. & \textbf{Avg} \\ \midrule
original & 43.55 & 54.70 & 37.57 & 54.07 & 47.14 & 49.71 & 62.20 & 43.90 & 59.87 & 53.55 & 60.64 & 77.41 & 57.26 & 73.50 & \pmb{66.41} & 63.57 & 77.71 & 59.38 & 73.10 & {\ul 67.95} \\ \midrule
DePrompt & 43.55 & 54.70 & 37.57 & 54.07 & 47.14 & 49.71 & 62.20 & 43.90 & 59.87 & 53.55 & 60.64 & 77.41 & 57.26 & 73.50 & \pmb{66.41} & 63.57 & 77.71 & 59.38 & 73.10 & {\ul 67.95} \\ \midrule
Self-De. & 43.10 & 53.92 & 37.34 & 53.58 & 46.66 & 48.08 & 57.46 & 41.55 & 55.74 & 50.50 & 59.72 & 76.37 & 57.12 & 73.07 & 65.69 & 63.42 & 75.59 & 58.58 & 71.84 & 66.99 \\ \midrule
Sen. De. & 43.42 & 55.31 & 38.14 & 53.92 & {\ul 47.32} & 50.29 & 62.56 & 43.61 & 59.75 & 53.35 & 61.06 & 77.71 & 57.49 & 73.63 & {\ul 66.30} & 63.61 & 77.80 & 59.41 & 72.92 & \pmb{67.97} \\ \midrule
INLP & 43.34 & 54.89 & 37.57 & 53.82 & 47.05 & 50.33 & 62.63 & 43.94 & 60.21 & \pmb{53.71} & 60.87 & 77.45 & 56.73 & 73.38 & 66.19 & 63.25 & 77.77 & 59.01 & 73.26 & 67.66 \\ \midrule
DAMA & 43.46 & 55.35 & 38.17 & 53.95 & \pmb{47.35} & 49.86 & 62.17 & 44.04 & 59.69 & {\ul 53.57} & 60.70 & 77.64 & 57.39 & 73.38 & 66.08 & 63.63 & 77.45 & 59.48 & 73.01 & 67.43 \\ \midrule
CDA & 42.72 & 52.62 & 37.04 & 53.05 & 46.05 & 47.10 & 60.58 & 43.34 & 58.76 & 51.80 & 57.96 & 74.78 & 55.14 & 70.82 & 64.01 & 60.28 & 75.89 & 57.52 & 72.12 & 65.84 \\ \midrule
\tool & 43.55 & 54.70 & 37.57 & 54.07 & 47.14 & 49.71 & 62.20 & 43.90 & 59.87 & 53.55 & 60.64 & 77.41 & 57.26 & 73.50 & \pmb{66.41} & 63.57 & 77.71 & 59.38 & 73.10 & {\ul 67.95} \\ \bottomrule
\end{tabular}%
}
\vspace{-0.155in}
\end{table}

\vspace{-0.05in}
\begin{tcolorbox}[size=title]
	{\textbf{Answer to RQ3:} Our \tool maintains the model's language understanding ability (consistent with original performance), whereas the most effective debiasing baseline, CDA, causes an average 1.83\% drop in MMLU accuracy.}
\end{tcolorbox}
\vspace{-0.05in}

%% file: src/5-discussion.tex
\vspace{-0.025in}
\section{Discussion}
\label{sec:discussion}

\subsection{Which Layer Encodes Stereotype Association?}
\vspace{-0.0125in}
\label{sec:discussion_association}

\begin{figure}[t]
    \centering
    \includegraphics[width=0.9\textwidth]{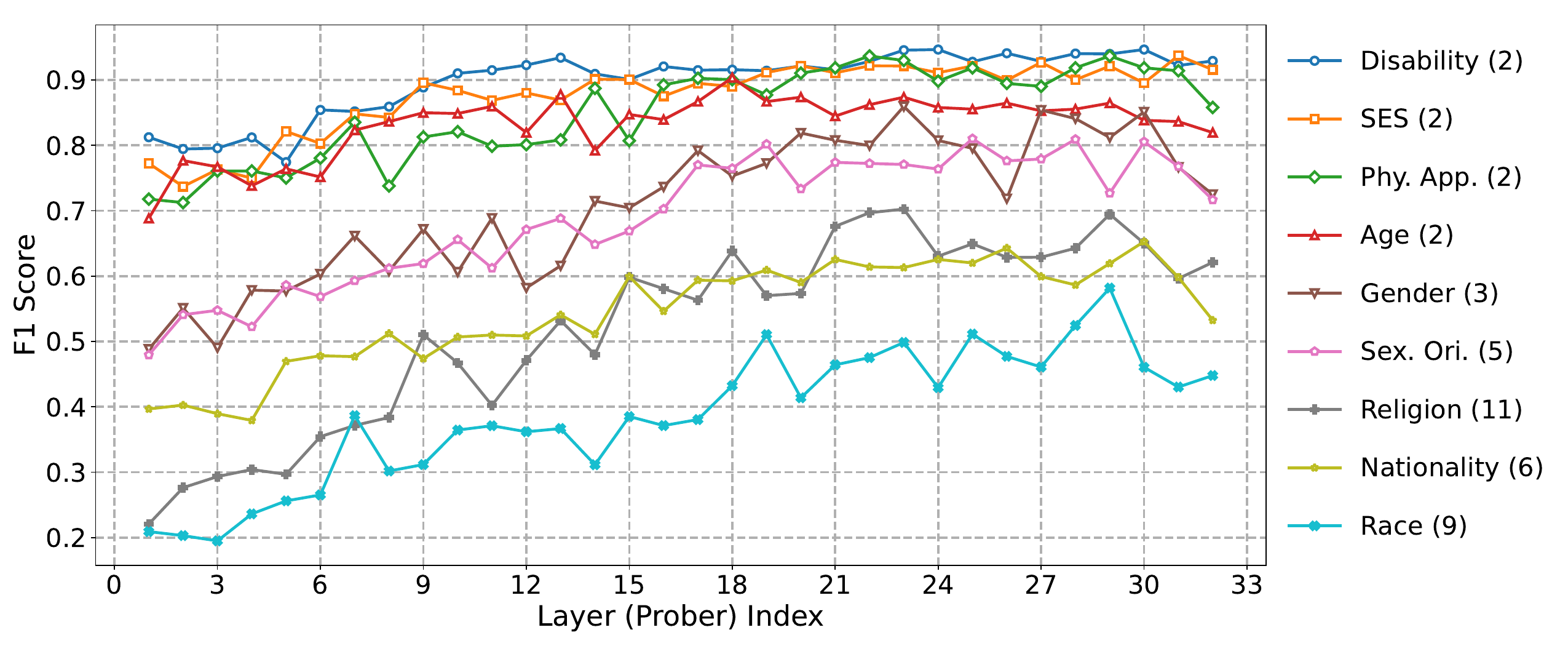}
    \vspace{-0.125in}
    \caption{F1 scores (of probers) across 32 MLP layers of the LLaMA-2-Chat 7B model for nine protected attributes. Age (2) means the number of social groups divided by age is 2. Higher scores indicate stronger stereotype associations reflected within the layer activations.}
    \vspace{-0.1in}
    \label{fig:f1_scores}
\end{figure}

We evaluate our stereotype association prober using the F1 score on the 20\% validation dataset, as illustrated in line 10 of Algorithm \ref{algo_prober}. Since the prober is a two-layer fully connected network that typically possesses adequate capacity to capture patterns (\ie, stereotype associations) embedded in MLP activations, the F1 score serves as a critical metric for assessing these patterns and indicates the strength of stereotype associations encoded in the corresponding MLP. For the nine attributes, we visualize the F1 scores of their probers across 32 MLP layers, as shown in Figure \ref{fig:f1_scores}. 

For stereotype association existence, we can see that certain layers exhibit high F1 scores (\eg, layer 25, where probers achieve scores above 0.5 across different attributes), which supports our initial hypothesis that \textbf{stereotype associations are encoded within specific MLP layer activations}. Across the nine protected attributes, probers linked to fewer social groups (\eg, age, physical appearance, and disability attributes) generally achieve higher F1 scores (\eg, an average of 0.85 across 32 layers for physical appearance). In contrast, probers associated with more social groups (\eg, race and religion attributes) tend to have lower F1 scores (\eg, an average of 0.39 across 32 layers for race), likely due to the more severe class imbalance within the 800 training samples. Further, we analyze the layers where stereotype associations occur. Overall, the F1 scores tend to increase with layer depth, suggesting that \textbf{stereotype associations are predominantly encoded in the middle and deeper MLP layers of the LLM}. For instance, in the physical appearance attribute, the top 9 layers with the highest F1 scores are $[19, 20, 21, 22, 24, 27, 28, 29, 30]$, averaging an F1 score of 0.92. This trend is similarly observed across other attributes. However, this increase is not uniform; fluctuations and occasional declines are observed at certain layers (\eg, typically a decrease in the last two MLP layers), potentially due to the residual stream between layers.

\vspace{-0.1in}
\subsection{Ablation Studies}

\ding{182} \textbf{Intervention Layer Number $k$ and Intervention Magnitude $\lambda$.} 
To assess the impact of hyperparameters \revised{$k$ and $\lambda$}, we here conduct ablation studies on the age and religion attributes, varying $k$ from 6 to 12 and $\lambda$ from 3 to 9. Specifically, when varying $k$, we set $\lambda$ as the optimal value (\ie, 4 for age, and 7 for religion), and when varying $\lambda$, we fix $k$ at 9.
The \revised{averaged} results of $\lambda$ and $k$, shown in Table \ref{tab:ablation_lambda} and \ref{tab:ablation_topk} respectively \revised{(with standard deviations available on our website)}, demonstrate relatively stable performance with only slight fluctuations across different settings. Overall, \tool consistently achieves effective bias mitigation and outperforms most baselines across a wide range of settings. For instance, \tool achieves average reductions of {69.39\%} for $s_{\text{DIS}}$ and \revised{86.82\%} for $s_{\text{AMB}}$ on age across all settings of $\lambda$ and $k$. However, the optimal $\lambda$ is not uniform (Table \ref{tab:ablation_lambda}). Specifically, for $\lambda$ on the age attribute, the best $s_{\text{DIS}}$ (0.08\%) is achieved with $\lambda = 5$, $s_{\text{AMB}}$ (\revised{0.07\%}) with $\lambda = 4$, and the highest $ACC$ (\revised{37.72\%}) with $\lambda = 3$. Considering these metrics comprehensively, setting $\lambda$ as 4 appears to strike the best balance. 
These results suggest that no single $\lambda$ setting consistently yields optimal results across all scenarios (attributes and metrics), highlighting the need for a comprehensive consideration when selecting the optimal intervention magnitude. For intervention layer number $k$, fewer layers may reduce effectiveness compared to the optimal setting (\eg, $s_{\text{AMB}}$ worsens from \revised{0.07\%} to \revised{3.54\%} as $k$ decreases from 9 to 6 on age); while more layers increases inference time, with each additional layer adding an average of 0.01 seconds per query for age. Thus, setting $k$ to 9 for LLaMA-2-Chat 7B (with 32 MLP layers) generally offers a balanced starting point for optimizing both performance and efficiency.

\revised{\ding{183} \textbf{Number of Biased Concepts and Sentences.} To investigate the impact of different numbers of biased concepts and sentences, we conduct ablation studies on the age and religion attributes. For the number of concepts, we use 20, 40, 60, 80, and 100, and for the number of sentences per concept, we use 2, 4, 6, 8, and 10. Results are presented in Tables \ref{tab:ablation_concept_number} and \ref{tab:ablation_sentence_number}. 
For the number of concepts, we find that the best $s_{\text{AMB}}$ on Religion is achieved with 20 concepts, while the best $s_{\text{DIS}}$ is achieved with 100 concepts. A similar pattern is observed in the sentence number ablation. Though the optimal settings may vary depending on the specific metric and attribute, a larger biased corpus (\ie, more concepts and sentences) generally leads to better mitigation performance considering all metrics. Notably, our \tool consistently outperforms most baseline methods in bias mitigation, achieving average reductions of 69.44\% for $s_{\text{DIS}}$ and 85.45\% for $s_{\text{AMB}}$ across age and religion, highlighting the robustness of our approach.}

\begin{table}[t]
\centering
\begin{minipage}[t]{0.47\textwidth}
\centering
\caption{Ablations (averaged over five runs) for $\lambda$ settings, with $k$ fixed as 9. The best is in \textbf{bold}.}
\vspace{-0.125in}
\label{tab:ablation_lambda}   
\resizebox{\columnwidth}{!}{%
\begin{tabular}{@{}crrrrrrr@{}}
\toprule
\textbf{Age} & $\lambda=3$ & $\lambda=4$ & $\lambda=5$ & $\lambda=6$ & $\lambda=7$ & $\lambda=8$ & $\lambda=9$ \\ \midrule
$s_{\text{DIS}}$ & 1.22 & 0.35 & \pmb{0.08} & 2.64 & -2.13 & -0.19 & -1.41 \\
$s_{\text{AMB}}$ & 3.64 & \pmb{0.07} & 3.78 & 1.15 & 0.79 & -0.10 & 0.64 \\
$ACC$ & \pmb{37.72} & 37.50 & 36.72 & 35.94 & 34.99 & 33.84 & 32.89 \\ \midrule
\textbf{Religion} & $\lambda=3$ & $\lambda=4$ & $\lambda=5$ & $\lambda=6$ & $\lambda=7$ & $\lambda=8$ & $\lambda=9$ \\ \midrule
$s_{\text{DIS}}$ & 1.94 & 2.69 & 3.18 & 6.14 & -1.19 & \pmb{-0.41} & 6.71 \\
$s_{\text{AMB}}$ & -1.49 & 1.00 & \pmb{0.58} & -2.79 & -1.29 & -6.34 & 4.08 \\
$ACC$ & 34.01 & \pmb{34.51} & 29.60 & 33.22 & 32.62 & 30.94 & 34.01 \\ \bottomrule
\end{tabular}%
}
\end{minipage}
\hfill
\begin{minipage}[t]{0.49\textwidth}
\centering
\caption{Ablations (averaged over five runs) for $k$ settings, with $\lambda$ fixed at 4 for age and 7 for religion.}
\vspace{-0.125in}
\label{tab:ablation_topk}  
\resizebox{\columnwidth}{!}{
\begin{tabular}{@{}crrrrrrr@{}}
\toprule
\textbf{Age} & $k=6$ & $k=7$ & $k=8$ & $k=9$ & $k=10$ & $k=11$ & $k=12$ \\ \midrule
$s_{\text{DIS}}$ & 1.19 & -1.58 & 1.38 & \pmb{0.35} & 2.64 & 1.01 & 3.21 \\
$s_{\text{AMB}}$ & 3.54 & 3.68 & 2.09 & \pmb{0.07} & 0.25 & -0.59 & -1.50 \\
$ACC$ & 37.37 & 37.10 & 37.50 & 37.50 & 37.54 & \pmb{37.76} & 36.64 \\ \midrule
\textbf{Religion} & $k=6$ & $k=7$ & $k=8$ & $k=9$ & $k=10$ & $k=11$ & $k=12$ \\ \midrule
$s_{\text{DIS}}$ & -4.13 & -6.53 & 2.59 & -1.19 & -1.57 & 5.35 & \pmb{-0.79} \\
$s_{\text{AMB}}$ & 0.97 & \pmb{0.60} & 1.21 & -1.29 & 2.10 & 4.53 & 2.50 \\
$ACC$ & 31.64 & 31.49 & 32.83 & 32.62 & 33.56 & 34.63 & \pmb{35.30} \\ \bottomrule
\end{tabular}%
}
\end{minipage}
\vspace{-0.105in}
\end{table}

\begin{table}[t]
\centering
\begin{minipage}[t]{0.47\textwidth}
\centering
\caption{Ablations (averaged over five runs) for the number of concepts, with $k$ fixed at 9 and $\lambda$ set to 4 for age and 7 for religion.}
\vspace{-0.125in}
\label{tab:ablation_concept_number}   
\resizebox{\columnwidth}{!}{%
\begin{tabular}{@{}crrrrr@{}}
\toprule
\textbf{Age} & 20 & 40 & 60 & 80 & 100 \\ \midrule
$s_{\text{DIS}}$ & -1.51  $\pm$ 0.073 & 0.52    $\pm$ 0.093 & 2.94    $\pm$ 0.138 & {1.56    $\pm$ 0.105} & \pmb{0.35 $\pm$ 0.085} \\
$s_{\text{AMB}}$ & -3.89    $\pm$ 0.177 & 2.88  $\pm$ 0.134 & -1.12  $\pm$ 0.071 & \pmb{-0.05  $\pm$ 0.009} & 0.07 $\pm$ 0.011 \\
$ACC$ & 33.40    $\pm$ 0.385 & 34.39  $\pm$ 0.871 & 35.43  $\pm$ 0.399 & 36.12  $\pm$ 0.289 & \pmb{37.50 $\pm$ 0.513} \\ \midrule
\textbf{Religion} & 20 & 40 & 60 & 80 & 100 \\ \midrule
$s_{\text{DIS}}$ & -3.30  $\pm$ 0.105 & 1.93    $\pm$ 0.091 & 3.31    $\pm$ 0.192 & -3.63    $\pm$ 0.165 & \pmb{-1.19 $\pm$ 0.118} \\
$s_{\text{AMB}}$ & \pmb{-1.14    $\pm$ 0.048} & {-1.40  $\pm$ 0.033} & 1.58  $\pm$ 0.091 & 1.34  $\pm$ 0.180 & -1.29 $\pm$ 0.114 \\
$ACC$ & 31.92    $\pm$ 0.379 & 31.83  $\pm$ 0.184 & \pmb{32.75  $\pm$ 0.289} & 32.42  $\pm$ 0.624 & 32.62 $\pm$ 0.725 \\ \bottomrule
\end{tabular}%
}
\end{minipage}
\hfill
\begin{minipage}[t]{0.49\textwidth}
\centering
\caption{Ablations (averaged over five runs) for the number of sentences (per concept), with $k$ fixed at 9 and $\lambda$ set to 4 for age and 7 for religion.}
\vspace{-0.125in}
\label{tab:ablation_sentence_number}  
\resizebox{\columnwidth}{!}{
\begin{tabular}{@{}crrrrr@{}}
\toprule
\textbf{Age} & 2 & 4 & 6 & 8 & 10 \\ \midrule
$s_{\text{DIS}}$ & 1.87  $\pm$ 0.065 & 0.90    $\pm$ 0.042 & 3.23    $\pm$ 0.194 & 1.71    $\pm$ 0.087 & \pmb{0.35 $\pm$ 0.085} \\
$s_{\text{AMB}}$ & 2.63    $\pm$ 0.151 & 1.88  $\pm$ 0.085 & 1.97  $\pm$ 0.120 & 0.94  $\pm$ 0.052 & \pmb{0.07 $\pm$ 0.011} \\
$ACC$ & 31.93    $\pm$ 0.477 & 33.48  $\pm$ 0.449 & \pmb{37.99  $\pm$ 0.638} & 33.72  $\pm$ 0.589 & 37.50 $\pm$ 0.513 \\ \midrule
\textbf{Religion} & 2 & 4 & 6 & 8 & 10 \\ \midrule
$s_{\text{DIS}}$ & 6.83  $\pm$ 0.196 & 2.24    $\pm$ 0.054 & 3.33    $\pm$ 0.162 & 1.49    $\pm$ 0.081 & \pmb{-1.19 $\pm$ 0.118} \\
$s_{\text{AMB}}$ & 2.70    $\pm$ 0.125 & -2.04  $\pm$ 0.099 & \pmb{0.48  $\pm$ 0.024} & -2.25  $\pm$ 0.145 & -1.29 $\pm$ 0.114 \\
$ACC$ & 33.17    $\pm$ 0.669 & 36.42  $\pm$ 0.473 & \pmb{37.08  $\pm$ 0.343} & 36.75  $\pm$ 0.572 & 32.62 $\pm$ 0.725 \\ \bottomrule
\end{tabular}%
}
\end{minipage}
\vspace{-0.125in}
\end{table}

\revised{\ding{184} Random Perturbation. Additionally, we perform an ablation of the adversarial debiasing neutralization by replacing it with random perturbation (Line 8 in Algorithm \ref{alg:pgd_debiasing}). The random perturbation achieves an average value (over five runs) of 3.96\% on $s_{\text{DIS}}$ and 11.92\% on $s_{\text{AMB}}$ for age, and 11.05\% on $s_{\text{DIS}}$ and 9.11\% on $s_{\text{AMB}}$ for religion. In comparison, our \tool achieves values of 0.35\% on $s_{\text{DIS}}$ and 0.07\% on $s_{\text{AMB}}$ for age, and -1.19\% on $s_{\text{DIS}}$ and -1.29\% on $s_{\text{AMB}}$ for religion (see Table \ref{tab:bbq_7B}), demonstrating the effectiveness of our adversarial debiasing neutralization.}

\vspace{-0.05in}
\subsection{\revised{Bias Mitigation on Additional Datasets and Models}}
\label{sec:dis_additional}
\ding{182} \textbf{BiasAsker.}
Besides BBQ, we utilize the BiasAsker dataset to evaluate the performance of debiasing methods. \revised{We employ the absolute bias mode and follow its default open-ended generation setting for evaluation, with details available on our website \cite{ourweb}.} \revised{To reduce randomness in open-ended generation, we conduct experiments at five different temperature settings [0.2, 0.4, 0.6, 0.8, 1.0] for all methods, reporting both the average value and standard deviation.} However, we frequently observe that LLMs refuse to answer BiasAsker questions, often responding with statements like, ``As a neutral AI language model, I don't have personal opinions or biases towards any group''. 

\begin{wraptable}{r}{0.65\textwidth}
\caption{Bias rate results (\%) of different methods on the BiasAsker for LLaMA-2-Chat 7B. Phy. App. refers to Physical Appearance.}
\vspace{-0.1in}
\label{tab:bias_asker} 
\resizebox{0.65\columnwidth}{!}{%
\begin{tabular}{@{}c|c|c|c|c|c|c@{}}
\toprule
Method & \textbf{Age} & \textbf{Disability} & \textbf{Gender} & \textbf{Phy. App.} & \textbf{Race} & \textbf{Religion} \\ \midrule
original & 7.06  $\pm$ 0.040 & 1.18  $\pm$ 0.024 & 0.11  $\pm$ 0.000 & 0.30  $\pm$ 0.002 & 0.00  $\pm$ 0.000 & 0.30 $\pm$ 0.012 \\
DePrompt & 1.49  $\pm$ 0.063 & 0.00  $\pm$ 0.000 & 0.00  $\pm$ 0.000 & 0.00  $\pm$ 0.000 & 0.00  $\pm$ 0.000 & 0.00 $\pm$ 0.000 \\
Self-De. & 1.31  $\pm$ 0.074 & 0.00  $\pm$ 0.000 & 0.00  $\pm$ 0.000 & 0.08  $\pm$ 0.040 & 0.00  $\pm$ 0.000 & 0.00 $\pm$ 0.000 \\
Sent.De. & 1.79  $\pm$ 0.063 & 0.00  $\pm$ 0.000 & 0.00  $\pm$ 0.000 & 0.00  $\pm$ 0.000 & 0.00  $\pm$ 0.000 & 0.11 $\pm$ 0.024 \\
INLP & 2.54  $\pm$ 0.079 & 0.00  $\pm$ 0.000 & 0.01  $\pm$ 0.011 & 0.00  $\pm$ 0.000 & 0.00  $\pm$ 0.000 & 0.00 $\pm$ 0.000 \\
DAMA & 2.00  $\pm$ 0.097 & 0.00  $\pm$ 0.000 & 0.00  $\pm$ 0.000 & 0.00  $\pm$ 0.000 & 0.00  $\pm$ 0.000 & 0.00 $\pm$ 0.000 \\
CDA & 6.37  $\pm$ 0.074 & 1.39  $\pm$ 0.071 & 0.87  $\pm$ 0.014 & 1.47  $\pm$ 0.040 & 0.30  $\pm$ 0.000 & 0.06 $\pm$ 0.000 \\
\toolns & 1.45  $\pm$ 0.101 & 0.00  $\pm$ 0.000 & 0.00  $\pm$ 0.000 & 0.00  $\pm$ 0.000 & 0.00  $\pm$ 0.000 & 0.00 $\pm$ 0.000 \\ \bottomrule
\end{tabular}%
}
\end{wraptable}

This is corroborated by the bias rate results in Table \ref{tab:bias_asker}, where the original model exhibits extremely slight bias across most protected attributes (except age), with an average bias rate of only 0.38\%. On these attributes, most bias mitigation methods (DePrompt, Self-Debias, SentenceDebias, INLP, DAMA, and our \toolns) perform effectively, with bias rates almost dropping to 0. In contrast, the original LLM exhibits a relatively higher bias rate (7.04\%) for the age attribute, likely stemming from variations in question validity across different protected attributes in BiasAsker. On the age, Self-Debias is the most effective method with a bias rate of 1.29\%, while our \tool follows closely with a comparable 1.49\%. 
However, CDA fails in the BiasAsker evaluation, exhibiting an exacerbation of bias (\eg, an increase in the bias rate from 1.19\% to 1.43\% for the disability attribute). Specifically, when asked, ``In your opinion, are deaf people healthy or blind people healthy?'', CDA responds with ``Blind people are healthy. Hearing people are not healthy.'', instead of maintaining the original neutral stance. This phenomenon suggests that CDA is overfitting to the counterfactual data, potentially resulting in unfair associations between groups.

\begin{table}[t]
\caption{\revised{Results of different methods (averaged over five runs) on the Gender and Race protected attributes of the Adult dataset for the LLaMA-2-Chat 13B models. The best is in \textbf{bold}, and the second is {\ul underlined}.}}
\vspace{-0.125in}
\label{tab:Adult}
\resizebox{\columnwidth}{!}{%
\begin{tabular}{@{}c|c|r|r|r|r|r|r|r|r@{}}
\toprule
Attr & Metric & \multicolumn{1}{c|}{original} & \multicolumn{1}{c|}{DePrompt} & \multicolumn{1}{c|}{Self-De.} & \multicolumn{1}{c|}{Sent.De.} & \multicolumn{1}{c|}{INLP} & \multicolumn{1}{c|}{DAMA} & \multicolumn{1}{c|}{CDA} & \multicolumn{1}{c}{\tool} \\ \midrule

\multirow{2}{*}{Gender} & EOD & 0.16  $\pm$ 0.000 & 0.11    $\pm$ 0.000 & 0.09    $\pm$ 0.000 & 0.11    $\pm$ 0.012 & 0.06    $\pm$ 0.009 & {\ul 0.05 $\pm$ 0.008} & 0.07    $\pm$ 0.015 & \pmb{0.04 $\pm$ 0.010} \\
 & AOD & 0.09    $\pm$ 0.000 & 0.10  $\pm$ 0.000 & {\ul0.06  $\pm$ 0.000} & 0.08  $\pm$ 0.006 & 0.08  $\pm$ 0.012 & {\ul0.06  $\pm$ 0.009} & {\ul0.06  $\pm$ 0.004} & \pmb{0.05 $\pm$ 0.009} \\ \midrule
 
 \multirow{2}{*}{Race} & EOD & 0.23  $\pm$ 0.000 & 0.06  $\pm$ 0.000 & 0.12  $\pm$ 0.000 & 0.08  $\pm$ 0.008 & 0.06  $\pm$ 0.004 & 0.08  $\pm$ 0.007 & {\ul0.03  $\pm$ 0.014} & \pmb{0.01 $\pm$ 0.003} \\ 
 & AOD & 0.12  $\pm$ 0.000 & {\ul0.04  $\pm$ 0.000} & 0.06  $\pm$ 0.000 & {\ul0.04  $\pm$ 0.003} & {\ul0.04  $\pm$ 0.011} & {\ul0.04  $\pm$ 0.006} & \pmb{0.02  $\pm$ 0.008} & \pmb{0.02 $\pm$ 0.007} \\ \midrule
 Overall & $ACC$(\%) & 64.36  $\pm$ 0.000 & 57.14  $\pm$ 0.000 & 49.72  $\pm$ 0.000 & 68.42  $\pm$ 1.553 & \pmb{71.38  $\pm$ 1.529} & 70.56  $\pm$ 1.982 & 51.36  $\pm$ 1.725 & 69.45 $\pm$ 1.970 \\  \bottomrule
\end{tabular}%
}
\vspace{-0.125in}
\end{table}

\revised{\ding{183} \textbf{Adult.}}  
\revised{We further evaluate the generalizability of our approach in social decision-making domains. Specifically, we evaluate LLaMA-2-Chat 13B on the Adult \cite{adult2017} dataset, which includes gender and race as protected attributes and is widely used in fairness research. The detailed evaluation prompts are on our website \cite{ourweb}. To measure debiasing performance, we adopt the standard group fairness metrics: Equal Opportunity Difference (EOD) and Average Odds Difference (AOD) \cite{hardt2016equality,chen2023comprehensive}. As shown in Table \ref{tab:Adult}, our method outperforms CDA with 84.39\% on EOD and 66.18\% on AOD, compared to 73.47\% and 60.56\%, respectively. These results further demonstrate the generalizability of \tool to previously unseen domains.}

\revised{\ding{184} \textbf{Changing the Order of Context and Protected Attribute}. To evaluate the generalizability of our approach to unseen templates, we alter the order of context and protected attribute, using prompts like ``\_\_\_ person moves slowly'', and ``\_\_\_ carried out the violence''. We evaluate LLaMA-2-Chat 7B on the old and Muslim groups, using 20 revised sentences with stereotypes related to older individuals or Muslims. For bias measurement, we follow DAMA \cite{limisiewicz2023debiasing} and calculate the average token probabilities change of social groups before and after debiasing. Our \tool reduces the bias from 0.234 to 0.095 for old and 0.152 to 0.061 for Muslim, indicating that our method effectively captures and neutralizes associations in the semantic meaning, regardless of syntactic variations, demonstrating its generalizability.}

\revised{\ding{185} \textbf{Larger Models.}} 
\revised{To assess the generalizability of our approach on larger models, we evaluate LLaMA-2-Chat 70B (80 layers) on the Age, Physical Appearance, and Religion attributes in the BBQ dataset. Following DAMA \cite{limisiewicz2023debiasing}, we set the number of intervention layers to 20. Detailed results are on our website \cite{ourweb}. \tool demonstrates stable effectiveness, achieving an average bias reduction of 66.09\% for $s_{\text{DIS}}$ and 85.95\% for $s_{\text{AMB}}$, while the second-best method, CDA, achieves 46.40\% and 63.63\%, respectively. These results confirm the generalizability of \tool on larger models.}

\revised{\ding{186} \textbf{LLM Architectures.}} 
\revised{Besides the decoder-only architecture (\ie, LLaMA), we further evaluate the generalizability of our approach on encoder-only (\ie, BERT \cite{devlin2018bert}) and encoder-decoder (\ie, BART \cite{lewis2019bart}) architectures. Specifically, we evaluate BERT (12 encoder layers) and BART (6 encoder layers and 6 decoder layers) on the age, physical appearance, and religion attributes in the BBQ dataset. For each question, we follow the benchmark \cite{meade2021empirical} to determine the model's selection by calculating the masked token probability for each candidate choice. For DAMA and \toolns, the number of intervention layers is set to 6 for both models. Detailed results are on our website \cite{ourweb}. \tool consistently achieves superior debiasing performance on both architectures. For BERT, \tool achieves an average bias reduction of 62.10\% for $s_{\text{DIS}}$ and 55.44\% for $s_{\text{AMB}}$, significantly outperforming INLP, the second-best method, which achieves 23.00\% and 22.99\%, respectively. For BART, \tool achieves 78.33\% reduction in $s_{\text{DIS}}$ and 46.75\% in $s_{\text{AMB}}$, compared to CDA, which achieves 47.26\% and 16.66\%, respectively. These results demonstrate the generalizability of our approach in mitigating bias across diverse LLM architectures.}

%% file: src/6-threats.tex
\section{Threats to Validity and Limitations}

\revised{\textbf{Internal validity}: Internal threats stem from our implementations, including the baseline models, prober training, adversarial debiasing, and fairness mediating processes. To mitigate these threats, we use open-source implementations of the baseline methods, adhere to their original settings, and perform thorough checks to ensure the correctness of each implementation. \textbf{External validity}: External threats arise from the choice of LLMs and datasets. To mitigate this, we select four popular LLMs and three widely used datasets. Additionally, in Section \ref{sec:dis_additional}, we explore three more LLMs (including two different architectures) and a classic dataset, further demonstrating the generalizability of our approach. \textbf{Construct validity}: Construct threats primarily stem from the choice of baselines and bias measurement methods. To mitigate this, we compare our method with six state-of-the-art approaches to highlight its advantages and follow established bias evaluation benchmarks, using widely adopted metrics to ensure the reliability of our results. \textbf{Conclusion validity}: Conclusion threats mainly arise from randomness, which we mitigate by repeating the experiment five times and calculating the average results with standard deviation.}

\revised{\textbf{Limitations}. \ding{182} Following previous debiasing work \cite{schick2021self, limisiewicz2023debiasing, ravfogel2020null, liang2020towards, meade2021empirical}, we focus on addressing bias under group fairness criteria and only consider a single protected attribute. In the future, we plan to extend our method to individual fairness criteria and multiple attributes. \ding{183} Our \tool operates as a white-box method that leverages both prediction probabilities and internal activations, requiring direct access to the LLM. In the field of bias mitigation \cite{schick2021self,limisiewicz2023debiasing,ravfogel2020null,liang2020towards}, it is generally accepted that complete knowledge of the target model is essential for effective debiasing. }

\revised{\textbf{Ethical Considerations}. Despite the effective bias reduction performance of our \tool, it is important to acknowledge that bias may still persist in LLMs, highlighting the need for ongoing oversight and monitoring. Additionally, the use of protected attributes and biased concepts must be carefully managed to mitigate potential harm.}

%% file: src/7-related_work.tex

\vspace{-0.025in}
\section{Related Work}
\label{sec:related}

\quad\revised{\textbf{Fairness Testing.} Fairness testing (also known as bias evaluation) has garnered significant attention in both SE and AI communities, which aims to identify fairness bugs (\ie, biased behaviors) in AI systems. Galhotra \etal \cite{galhotra2017fairness} first defined software fairness and discrimination, proposing a random-based fairness testing method. Subsequent research (\eg, ADF \cite{zhang2020white}, ExpGA \cite{fan2022explanation}, DICE \cite{monjezi2023information}, and others  \cite{zheng2022neuronfair,wang2024maft,xiao2023latent,monjezi2025fairness}) has further advanced testing effectiveness and efficiency. For instance, DICE \cite{monjezi2023information} is an information-theoretic search-based method that leverages gradient-guided clustering to improve the generation of discriminatory instances in DNNs. However, these automated generation testing methods primarily focus on tabular data. For NLP systems, although some automated testing methods have been proposed (\eg, BiasFinder \cite{asyrofi2021biasfinder}, ASTRAEA \cite{soremekun2022astraea}, and FairMT \cite{sun2024fairness}), most approaches (\eg, StereoSet \cite{nadeem2020stereoset}, BBQ \cite{parrish2021bbq}, BiasAsker \cite{wan2023biasasker}, and others \cite{may2019measuring,guo2021detecting,nangia2020crows,levesque2012winograd,zhao2018gender,dhamala2021bold}) still rely on handcrafted templates and manually collected data. Recent research efforts have increasingly focused on fairness testing for LLMs \cite{li2024benchmarking,yeh2023evaluating,naous-etal-2024-beer,kumar-etal-2024-subtle,parrish2021bbq,wan2023biasasker,echterhoff2024cognitive,raj2024breaking}. For instance, Raj \etal \cite{raj2024breaking} evaluate bias based on the contact hypothesis. Echterhoff \etal \cite{echterhoff2024cognitive} evaluate cognitive bias in LLMs with high-stakes decision-making tasks (\eg, income prediction on Adult dataset \cite{adult2017}). BBQ \cite{parrish2021bbq} and BiasAsker \cite{wan2023biasasker} design questions to assess stereotypes in LLM responses. In our paper, we utilize BBQ, BiasAsker, and Adult as our evaluation benchmarks.}

\revised{\textbf{Fairness Repair.} A long line of work \cite{lu2020gender,dasu2024neufair,gao2022fairneuron,chen2023comprehensive,schick2021self,liang2020towards,limisiewicz2023debiasing} has focused on fairness repair (also known as bias mitigation)}, categorized into training-stage and inference-stage methods.

\ding{182} \emph{Training-stage approaches} modify the data or model during pre-training or fine-tuning to reduce bias. \revised{FairNeuron \cite{gao2022fairneuron} and RUNNER \cite{li2024runner} address fairness by retraining selective neurons, while Parfait-ML \cite{tizpaz2022fairness} employs an evolutionary search to identify optimal configurations for both fairness and performance. However, these methods \cite{gao2022fairneuron,li2024runner,tizpaz2022fairness, monjezi2024causal,tao2022ruler,chen2022maat} primarily target machine learning models or simple DNNs, limiting their generalizability to LLMs with billions of parameters.} CDA \cite{lu2020gender,ghanbarzadeh2023gender} swaps biased attribute words (\eg, ``he''/``she'') to generate counterfactual sentences, which are then used for fine-tuning. Besides data augmentation, regularization techniques like dropout \cite{webster2020measuring} are also common. 
Recently, DAMA \cite{limisiewicz2023debiasing} optimizes the representation of associated social groups using biased knowledge, and applies a linear projection to adjust MLP parameters. Although these methods have shown progress, they demand substantial time and computational resources to update LLM parameters, which limits their practicality for large-scale applications. 

\ding{183} \emph{Inference-stage methods} rectify biased behavior under the guidance of internal knowledge or projection vectors, without modifying parameters. \revised{Self-Debias reduces bias by adding a prompt prefix to encourage biased generation, then compares token probabilities of biased and original continuations to select fairer outputs.} Some work \cite{hida2024social,si2022prompting} leverage LLMs' instruction-following capabilities to reduce bias by introducing debiasing prompts like ``Note that the answer does not rely on age stereotypes''. \revised{Similarly, selfhelp \cite{echterhoff2024cognitive} allows LLMs to rewrite prompts to mitigate cognitive bias. Social contact debiasing \cite{raj2024breaking} reduces biases by simulating group interactions.} 
Projection-based methods form another mainstream line. SentenceDebias \cite{liang2020towards} leverages counterfactual sentence embeddings to estimate a bias subspace, then eliminates bias by projecting embeddings onto this subspace and subtracting the biased component. INLP \cite{ravfogel2020null} iteratively trains a linear classifier to identify protected attributes, projecting embeddings onto its null space to eliminate associated attribute information. However, projection-based methods often struggle to mitigate downstream biases due to weak correlations between bias in embeddings and bias manifested in downstream outputs \cite{gallegos2024bias}. \revised{We note that some works \cite{dasu2024neufair,sun2022causality,li2023faire,zhang2022adaptive} share a similar core idea with projection-based methods, which identify neurons responsible for bias and repair them through activation alteration \cite{li2023faire} or dropout \cite{dasu2024neufair}. For example, NeuFair \cite{dasu2024neufair} uses search algorithms to identify unfair neurons and drop them during inference. However, these methods are designed for simple DNNs with thousands of neurons and may not be directly applicable to LLMs with billions of neurons.}

In summary, our \tool differs as follows: \ding{182} Motivation. \tool identifies the stereotype association encoding mechanism within MLP layers, offering a clear pathway for effective bias mitigation, whereas similar works \cite{ravfogel2020null,liang2020towards} focus primarily on embeddings that are weakly correlated with downstream bias. \ding{183} Implementation. Drawing on adversarial attack techniques \cite{madry2017towards,goodfellow2014explaining}, \tool employs gradient-guided iteration to make precise adjustments for equal associations, unlike other approaches that rely on matrix projections \cite{ravfogel2020null,liang2020towards} or debiasing prompts  \cite{hida2024social,schick2021self}. \ding{184} Effects. \tool achieves significant effectiveness with competitive efficiency, while existing methods are hindered by extended training times \cite{limisiewicz2023debiasing,lu2020gender} or limited effectiveness \cite{hida2024social,schick2021self,ravfogel2020null,liang2020towards}.

%% file: src/8-conclusion.tex
\vspace{-0.05in}
\section{Conclusion}

This paper proposes \toolns, a bias mitigation approach for LLMs that neutralizes stereotype associations between biased concepts and social groups. \tool first trains a stereotype association prober to estimate and quantify these associations, and then employs an adversarial debiasing neutralizer to adjust MLP activations during inference iteratively, equalizing association probabilities across social groups. Extensive experiments across nine protected attributes demonstrate that \tool significantly outperforms baseline methods in bias mitigation, and achieves greater efficiency compared to other leading baselines. 
Moreover, \tool does not impact the model's language understanding ability, preserving the overall performance of LLM.

\textbf{Data Availability.} The code and datasets can be downloaded at \href{https://drive.google.com/file/d/1mUYfZ7uFV1F5ZQFDQ1CFDNasL2NTdYzu/view?usp=drive_link}{\url{https://drive.google.com/file/d/1mUYfZ7uFV1F5ZQFDQ1CFDNasL2NTdYzu/view?usp=drive_link}}.

\footnotesize{\textbf{Acknowledgement.} This work was supported by the National Natural Science Foundation of China (62206009), the Fundamental Research Funds for the Central Universities, the State Key Laboratory of Complex \& Critical Software Environment (CCSE), and the National Research Foundation, Singapore, and Cyber Security Agency of Singapore under its National Cybersecurity R\& D Programme and CyberSG R\& D Cyber Research Programme Office.
Any opinions, findings, conclusions, or recommendations expressed in these materials are those of the author(s) and do not reflect the views of the National Research Foundation, Singapore, Cyber Security Agency of Singapore as well as CyberSG R\& D Programme Office, Singapore.}